%% file: main.tex
\documentclass[a4paper,11pt]{article}
\pdfoutput=1 
\usepackage{jcappub} 
\usepackage[T1]{fontenc} 

\usepackage{url}
\usepackage{amssymb}
\newcommand{\xm}{$X_{\mathrm{max}}$ }
\newcommand\arcdeg{\mbox{$^\circ$}}%
\newcommand{\degree}{\arcdeg}

\title{A Commensal Radio-Only Cosmic Ray Detector at the Owens Valley Radio Observatory Long Wavelength Array}
\input{CR-instrument-paper-authors}

\abstract{The brief (10\,nanoseconds) transient radio emission from cosmic ray air showers carries key information about the energy and mass composition of high energy cosmic rays, but anthropogenic radio frequency interference has historically prevented radio-based cosmic ray studies from being carried out independently from other types of detectors.  We describe a cosmic ray detection system for the Owens Valley Radio Observatory Long Wavelength Array that searches for radio emission from cosmic ray air showers without relying on an external trigger, and runs alongside the other observing modes of the array.
 The \hbox{OVRO-LWA}, located in Eastern California, recently completed an expansion to 352 dual-polarization antennas and new signal processing infrastructure.  In order to detect cosmic rays in the presence of radio frequency interference (RFI), initial event classification and RFI rejection is performed on Field Programmable Gate Array boards, which each process a sampled voltage timeseries from both polarizations of a subarray of~32 antennas. Each board uses dedicated RFI veto antennas outside the air shower radio footprint to reject RFI events. We present the trigger design, RFI flagging strategy, and candidate cosmic rays.}   

\begin{document}
\maketitle
\flushbottom

\section{Introduction}
Several recent pieces of observational evidence suggest that the transition from Galactic to extragalactic sources of cosmic rays occurs somewhere between a few tens and a few thousands of PeV: gradual breaks in the cosmic ray energy spectrum around~1 and~100\,PeV and a flattening at a few thousand PeV (called the ``knee,'' the ``second knee,'' and the ``ankle,'' respectively); a shift in anisotropy around~8000\,\hbox{PeV}, pointing away from the Galactic plane \cite{Aab2017, Roulet2019}; and several shifts in composition across these energies \cite{Aab2016,Abbasi2019, Corstanje2021a}. Placing the shift at the lower end of this range creates challenges explaining certain features of the cosmic ray spectrum at high energies, but placing the shift at the high end of the range requires a second class (or classes) of Milky Way cosmic ray sources beyond supernova remnants \cite{Kachelrie2019,Thoudam2016,BeckerTjus2020}. 

In the energy range of the transition, the trajectories of charged cosmic ray nuclei traveling through the interstellar magnetic fields cannot point directly to their sources, making the energy-dependent shifts in cosmic ray mass composition a crucial observable.
Due to their low flux, high energy cosmic rays must be observed by studying the extensive air showers produced by their interaction with Earth's atmosphere. Mass-sensitive features of these air showers include the atmospheric column depth of the location of maximum particle density in the air shower, ~\xm, and the ratio of electron and muon secondary particles.  
The value of~\xm can be measured by observing optical fluorescence from the shower or the brief ($\sim 10\,\mathrm{ns}$), broadband (tens to hundreds of MHz) flashes of radio emission produced predominantly by the geomagnetic emission of high energy charged shower particles accelerating in Earth's magnetic field (with a lesser Askaryan radiation component due to  the charge excess at the shower front)\cite{Schroder2017}. 

The radio approach offers several advantages: radio antennas can instrument a large area at lower cost than particle detectors or fluorescence telescopes, and can operate with near full duty cycle in varying weather conditions. Furthermore, since electromagnetic interactions produce the radio emission, fewer model uncertainties affect the interpretation of the radio emission than in alternative techniques relying on the muon component of the showers. The amplitude of the pulse scales with the energy of the cosmic ray primary, and the shower radio emission is detectable against the background of Galactic synchrotron emission for cosmic rays with energies above a few tens of PeV.
In addition to these practical advantages, new radio observations of air showers by dense radio arrays have a key science motivation by enabling the application of recently proposed techniques\cite{schoorlemmer2021interferometry} that go beyond \xm to further reduce the model-dependent systematic uncertainties\cite{Buitink2021} that currently limit our understanding of cosmic ray composition, by measuring the 3D shapes of the cosmic ray air showers.

Thus far, however, the challenge of detecting radio emission from cosmic rays amid a background of human-generated radio frequency interference (RFI) has hampered radio techniques from reaching their full potential.  Radio-only triggers operate in extremely radio-quiet environments such as Greenland and Antarctica (e.g., \cite{RNO-G2025,ANITA2021}) for tau neutrino searches, but, in other environments, the experiments that have generated significant samples of radio-observed cosmic rays rely on particle detectors either for the initial trigger to read out data, or for classifying events as cosmic rays (e.g.\ \cite{Buitink2016, Pont2021}).  

The Owens Valley Radio Observatory Long Wavelength Array (OVRO-LWA,\cite{Monroe2020,Eastwood2019,anderson2019OVROLWA-transients}) has an antenna layout suitable to sample cosmic ray air shower footprints with unprecedented resolution, facilitating precision composition studies, and now has undergone a major upgrade.

To achieve the science goal of advancing the understanding of cosmic ray composition at the Galactic to extragalactic transition, the OVRO-LWA meets the following requirements:  hundreds of antennas within a square kilometer, an ability to run the cosmic ray system in parallel with other observing modes, radio-only triggering that operates amid RFI, a timing precision of better than 5 nanoseconds, and an ability to use Galactic synchrotron emission for calibration.
In this paper we present a standalone-radio cosmic ray detection system that utilizes the full array and runs in parallel with the other science modes of the array, in normal RFI conditions, and show how it meets these requirements.   Section~\ref{ovro} gives an overview of the \hbox{OVRO-LWA}.  Section~\ref{sec:trigger} presents the radio-only trigger, the event classification system, and the commissioning dataset, section~\ref{cal} describes calibration, and section~\ref{acceptance} characterizes the detector acceptance. Sections~\ref{sec:discussion} and \ref{sec:conclusion} discuss future directions and conclude. 

\section{The Owens Valley Radio Observatory Long Wavelength Array (OVRO-LWA)}\label{ovro}

Caltech's Owens Valley Radio Observatory (OVRO) lies in a somewhat sparsely populated region of Eastern California, in a narrow valley between two tall mountain ranges that each exceed 4000~meters above sea level. The Sierra Nevada to the west and the White Mountains to the east help block RFI from major cities, but RFI sources affecting the valley include power lines, air planes, and local activity. The Owens Valley Radio Observatory sits at the valley floor at an elevation of 1200~meters, well below \xm for the air showers we expect to observe. A 100\,PeV proton arriving from zenith with a typical \xm of 700\,g\,cm${}^{-2}$ will reach \xm at an altitude of roughly 3000~meters. 

\begin{figure}[tb]
    \centering
    \includegraphics[width=0.7\textwidth]{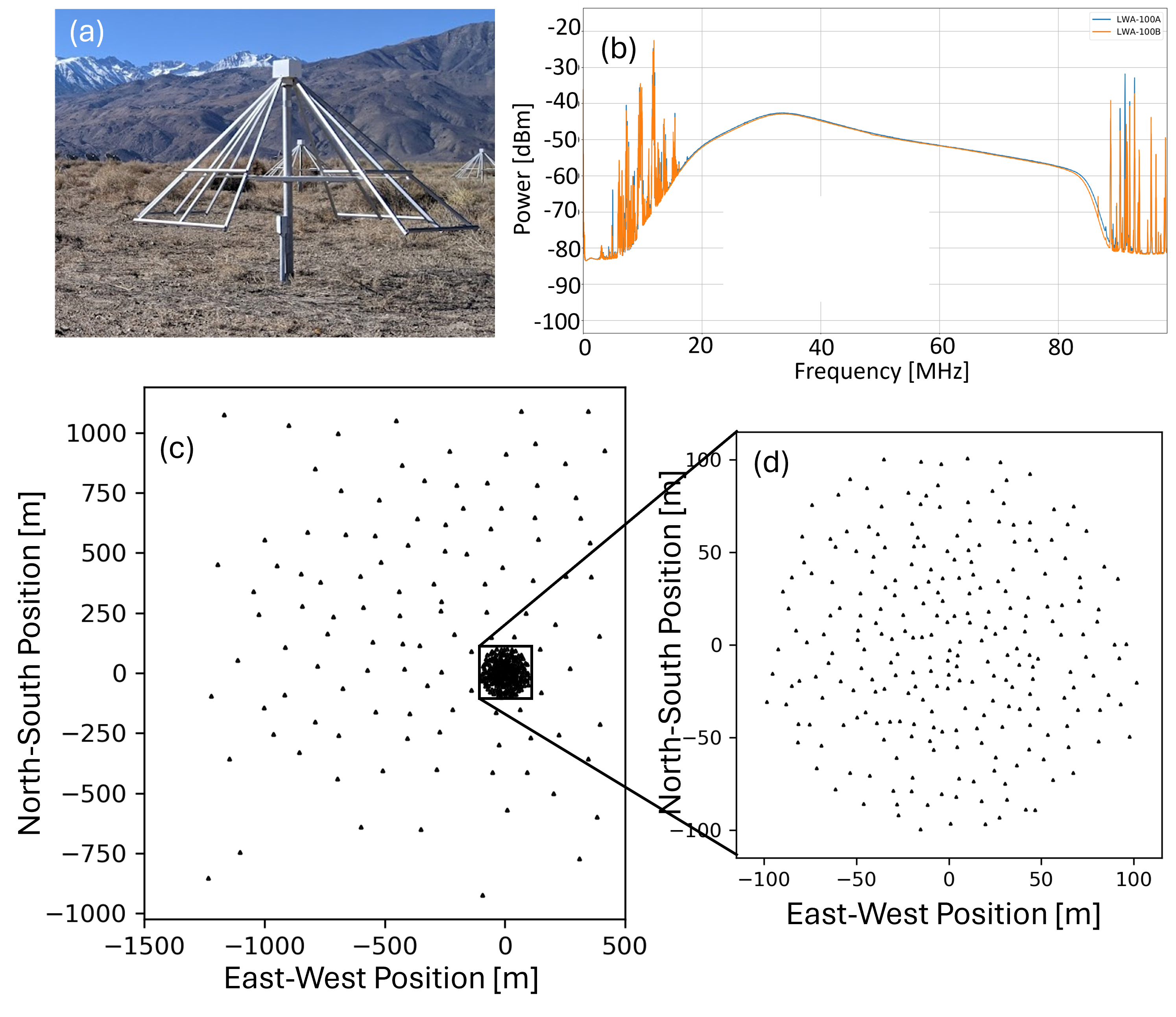}
    \caption{ (a)~OVRO-LWA antenna. Front-end electronics sit under the white square cover at the top of the antenna. (b)~Typical OVRO-LWA spectrum from both polarizations of one antenna stand (the same antenna shown in panel a), including the response of the antenna and the entire analog signal path, with the spectra from the East-West-aligned dipole and the North-South-aligned dipole in orange and blue, respectively. (c)~Layout of the 352 dipole pairs. (d)~Inset showing the layout of the dense core array. }
    \label{antenna}
\end{figure}

The OVRO-LWA (Figure \ref{antenna}) is an array of dual polarization dipoles built to study a wide range of phenomena including extrasolar space weather, the cosmic dawn epoch, solar flares, as well as now searching for high-energy cosmic rays. The OVRO-LWA was built in three stages. The first stage consisted of 256 dual-polarization dipole antennas in a dense array roughly 200\,m in diameter. A second stage added 32 antennas outside this dense core array \cite{Eastwood2019, Anderson2019LWAtransients}. Finally, from 2019 to 2023, the array underwent an upgrade funded by the NSF's MRI program that consisted of expanding from~288 to~352 antennas within a 2\,km extent, re-designing the analog electronics for better signal path isolation and better sensitivity to lower frequencies, and entirely replacing the digital electronics.

Monroe et al.~\cite{Monroe2020} demonstrated standalone-radio cosmic ray detection with the second-stage array OVRO-LWA in a short, dedicated experiment during an interval of relatively reduced \hbox{RFI}. They detected eight cosmic ray air showers consistent with 80--200\,PeV cosmic rays in 40\,hours. The new digital signal processing electronics in the upgraded array created the opportunity to build a cosmic ray detection system that will search for air showers alongside the other types of radio astronomy observations. 

The final layout of the upgraded OVRO-LWA consists of 256 antennas spaced irregularly in a dense core of~200~meters in diameter, surrounded by sparsely-spaced antennas out to a maximum separation of 2.4\,km (Figure~\ref{antenna}).    The OVRO-LWA array layout was primarily driven by requirements for interferometric imaging, but is suitable for detecting $\sim$100--1000\,PeV cosmic rays: most cosmic rays will be detected within the core of the array, whereas the sparser regions can detect inclined showers at higher energies, and the most distant antennas enable the RFI rejection strategy presented in section \ref{sec:fpgaveto}. The cosmic ray science case contributed to the decision to place additional antennas just outside of the dense core array (\cite{mythesis}). 

Panel~(a) of Figure \ref{antenna} shows an LWA antenna, and panel~(b) shows a typical power spectrum observed by both polarizations of a typical antenna.
Each antenna stand has one dipole aligned to geographic North-South and the other dipole aligned to geographic East-West. These antennas are not designed to distinguish horizontal and vertical polarization. The antenna's sensitivity is dominated by a diffuse background of Galactic synchrotron emission over two octaves in frequency, from~18 to~85\,MHz (see Section \ref{sec:efieldcal} for details on the Galactic emission in the 30--80\,MHz band used for the cosmic ray search). 

\begin{figure}
    \centering
    \includegraphics[ width=0.9\textwidth]{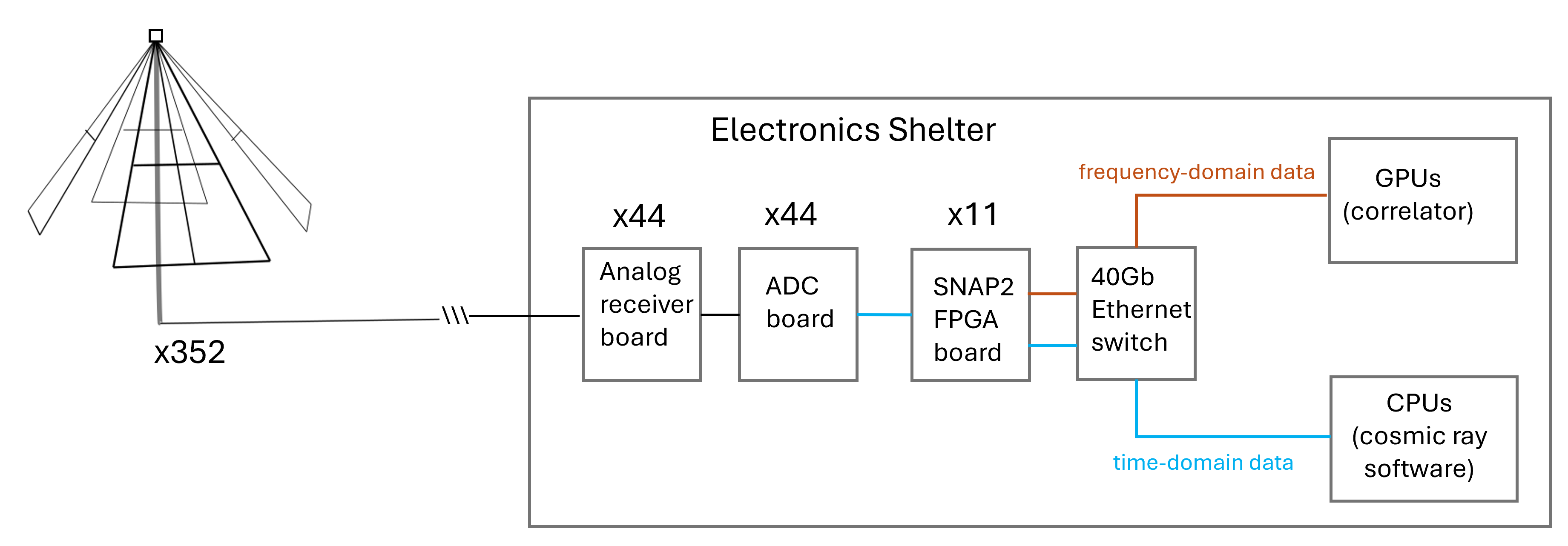}
    \caption{ Left: Simplified diagram of the OVRO-LWA signal path. Signals from 352 dual polarization antennas are transmitted to a central location, where 44 analog circuit boards filter 16 signals each, passing the signals to~44 digitization (analog-to-digital conversion [ADC]) boards, each of which samples 16 signals each. Eleven SNAP2 boards process the output of the ADCs, and the cosmic ray trigger logic runs on these boards. A 40\,Gb switch allows cosmic ray data to be transmitted to a separate computer from the correlator.} 
    \label{fig:lwa-signalpath}
\end{figure}

At the top of each antenna is a front-end electronics enclosure which houses a printed circuit board containing, for each of the two dipoles, a low pass filter (5th order Butterworth with 3dB point at 150 MHz), a low noise amplifier (37 dB of gain) and a bias tee that allows the one coaxial cable per dipole to both supply DC power to the low noise amplifier and carry the RF signal \cite{Hicks2009LWA}.
The analog signals from each antenna are transmitted to a central electronics shelter (Figure~\ref{fig:lwa-signalpath}) using coaxial cable for antennas in the dense core array. For the more distant antennas, a short length of coaxial cable links the antenna to a junction box where the RF signal modulates a laser driver to send the analog signal over fiber optics cable to the central electronics shelter.
The central electronics shelter itself is a modified shipping container. Additional radio-frequency shielding has been added to it to minimize the amount of self-interference generated by emissions radiating from the shelter and being received by the antennas.

Within the central electronics shelter, the signals are processed by a set of analog receiver (ARX) circuit boards, which contain several options for analog filters and attenuators, controllable from software.  The standard observing mode sets the high-pass cutoff at~18\,MHz and the low-pass cutoff at~85\,MHz. The attenuation settings are adjusted for each antenna with the goal of compensating for the different cable attenuations such that all ADC inputs receive similar power.
The signals pass from the ARX boards to be digitized by 10-bit analog-to-digital converters (ADCs) at a sample rate of 196\,MHz, producing a total data rate of 1.4\,Tbps across the array. \ref{appendix:timing} discusses the system for synchronizing the ADCs and timestamping data.

\begin{figure}[tb]
    \centering
    \includegraphics[ width=0.6\textwidth,trim={0cm 0cm 0cm 3cm},clip]{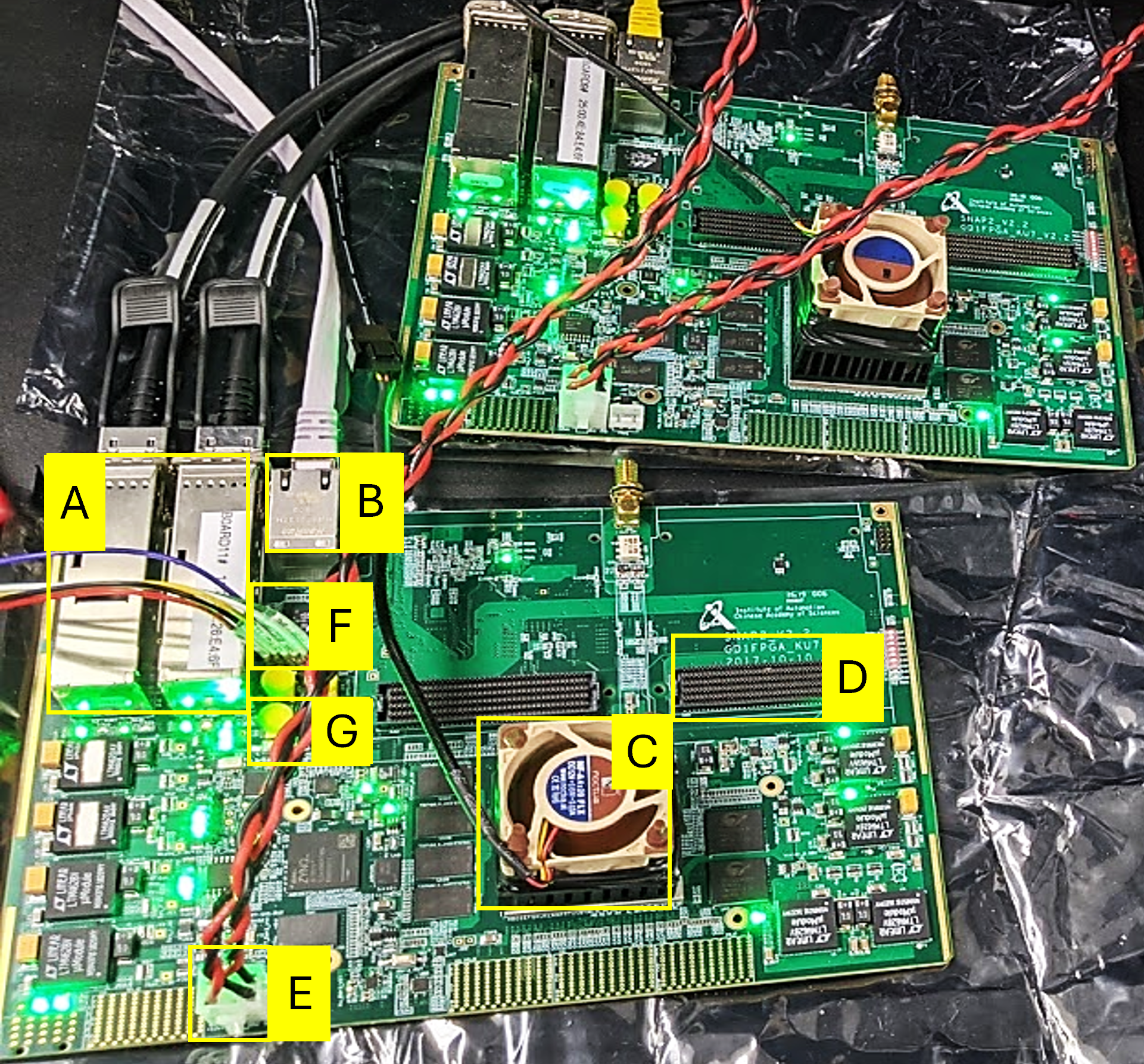}
    \caption{ SNAP2 boards with important features labeled: A)~40\,Gb QSFP+ Ethernet ports. B)~1\,Gb Ethernet ports. C)~The Xilinx Kintex Ultrascale FPGA is under this heat sink and fan--- the fan is used in the lab only, since in the OVRO-LWA cooled air flows through the entire electronics rack. D)~One of two FPGA-mezzanine-card (FMC) connectors. A stacked pair of ADC cards (not pictured) attaches to each FMC connector. E)~Power supply connector. F)~Pin header for programming the board via \hbox{JTAG}. After using JTAG to load an initial firmware configuration, subsequent re-programmings use the 1\,Gb Ethernet interface. G) SMA input for pulse-per-second (PPS) signal.} 
    \label{fig:snap2}
\end{figure}

The first stage of the cosmic ray search runs on 11 field programmable gate array (FPGA) boards, which are the second generation of the SNAP2,\footnote{\url{https://github.com/casper-astro/casper-hardware/blob/master/FPGA\_Hosts/SNAP2/README.md}} containing FPGAs, two 1\,Gb Ethernet ports, and four 40\,Gb QSFP+ Ethernet ports (Figure~\ref{fig:snap2}).
Each SNAP2 contains two FPGAs: a Xilinx Kintex Ultrascale and a smaller Xilinx Zynq. The larger Ultrascale runs the cosmic ray trigger, which operates in the time domain, in parallel with a filterbank that computes 4096 frequency-domain channels for use by the other observing modes. The Ultrascale additionally runs a Microblaze CPU that facilitates control of the FPGA. The smaller Zynq FPGA is used for tasks such as monitoring the SNAP2's power use. Each SNAP2 board processes signals from both polarizations of~32~antennas; four 16-channel ADC cards connect to each SNAP2. 

\section{Radio-only Cosmic Ray Detection}\label{sec:trigger}

 The OVRO-LWA's cosmic ray detection system has two main components.  First, search logic implemented on the FPGAs identifies events that are sufficiently cosmic-ray-like to trigger data readout (see also \cite{mythesis}). Second, these data are transmitted via Ethernet to computers that run the second stage of the system, classifying events as RFI or cosmic ray air shower candidates.  

Since the array upgrade replaced all the signal processing electronics, the new cosmic ray detection consists of all new firmware and software that builds on some of the techniques demonstrated by Monroe et al.~\cite{Monroe2020}, with some important differences.
The key differences in capability between this new system and that demonstration are operation in typical RFI conditions, operation in parallel with the other observing modes, and sampling shower footprints over a larger area.
These new capabilities are enabled by the new signal processing firmware and hardware that allows the cosmic ray search to use the full array and apply new RFI rejection methods.

\subsection{Coincident Pulse Search Logic}\label{sec:coincident}

\begin{figure}
    \centering
    \includegraphics[ width=0.9\textwidth]{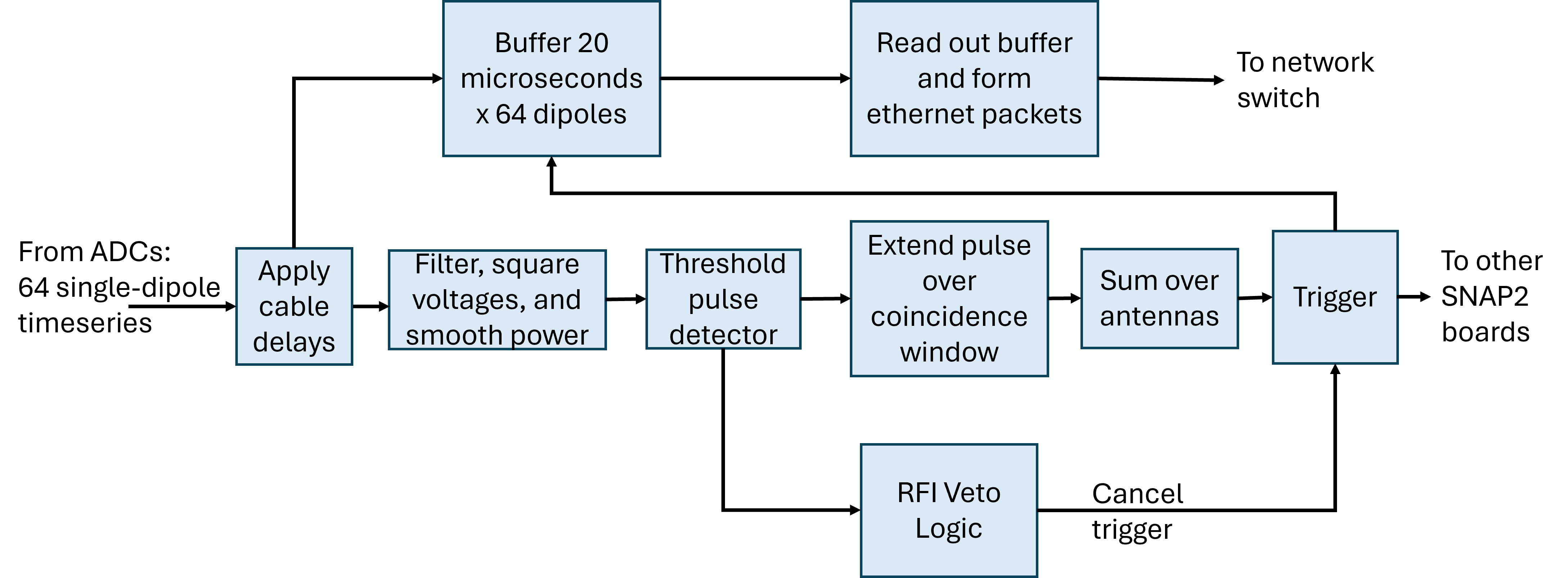}
    \caption{Flowchart summarizing the trigger logic for the cosmic ray radio-only trigger. This process is implemented in firmware running on the SNAP2 FGPA boards.}
    \label{fig:trigger}
\end{figure}

The total data rate output by all the ADCs is $~1.4$ Tbps: only snapshots of buffered data that meet a trigger condition can be read out to computers.   This subsection details the firmware logic that runs on the FPGAs on the SNAP2 boards, running alongside the filterbank firmware used by the other observing modes of the array.  Figure~\ref{fig:trigger} summarizes the trigger logic.

 Prior to the data entering the cosmic ray subsystem of the firmware, the ADC output for each antenna passes through delay logic to compensate for the relative differences in cable delays, due to the differing signal cable lengths for different antenna locations. The length and material of the cables and optical fiber determines the initial estimate of the delays used in the trigger. Section \ref{sec:cal.delay} describes the delay calibration process used later for data analysis. 

 The SNAP2s buffer 20\,$\mu$s of the time series output by the ADCs. The ADC sample rate is 196\,MHz, corresponding to a time resolution of 5.1\,ns. The buffer uses Block RAM (BRAM) implemented as a circular buffer that continuously overwrites old samples such that it always contains the most recent 20\,$\mu$s of data. Each SNAP2 handles 64 signals, from 32 dipole pairs. Within the firmware, every signal is treated as an individual antenna; knowledge of co-located polarization pairs becomes useful later in software processing.

A threshold-coincidence trigger operates in parallel with the buffer logic. This trigger logic first applies a time-domain finite impulse response (FIR) bandpass filter to the raw voltage timeseries. Although the analog electronics suppress the FM band as well as RFI below 30 MHz, some of these narrow-band RFI sources are so bright as to require additional suppression with the digital FIR filter. The FIR is a 24-tap equirripple filter providing a minimum of 20\,dB attenuation outside of the first nulls on either side of the passband, and additional suppression near the nulls. The filter is tuned to position a null to suppress a strong RFI source at 27\,MHz.
After filtering, the firmware squares the filtered voltages to produce a power timeseries, and smooths the power timeseries with a 4-tap moving window sum.  To summarize, this logic computes a smoothed power timeseries 
\begin{equation}
     p_i = \sum _{i-3} ^i x_i ^2
\end{equation}
where $x_i$ is the convolution of the raw ADC timeseries and the FIR coefficients. Next, threshold logic applied to the $p_i$ timeseries outputs a Boolean 1 whenever $p_i > p_{th}$, where $p_{th}$ is a power threshold that can be controlled from software commands to the FPGAs.

In order to trigger data readout, more than a threshold number of individual dipole signal chains processed by an individual FPGA must each satisfy $p_i > p_{th}$ within a window in time (coincidence window).  The length of the coincidence window is controllable from software and chosen as the maximum light travel time (in number of clock cycles) across the radius of the LWA core. The firmware implements counting the number of dipoles above threshold within the coincidence window by extending the Boolean ones output by the $p_i > p_{th}$ logic according to the coincidence window length and then summing across dipoles (see Figure 4).   
 If signals above the power threshold occur from more than a threshold number of dipoles, within the coincidence window, then the detector logic outputs a trigger signal, which can be cancelled by the RFI rejection system (\S\ref{sec:fpgaveto}). If the trigger is not cancelled by the RFI veto, then the FGPA stops writing new data to the buffer and transmits the buffered timeseries over Ethernet for the next stages of processing. The current operational settings require eight dipoles of any polarization to exceed the threshold to trigger, and three to veto (\S\ref{sec:fpgaveto}).

\begin{figure}[tb]
    \centering
    \includegraphics[ width=0.7\textwidth]{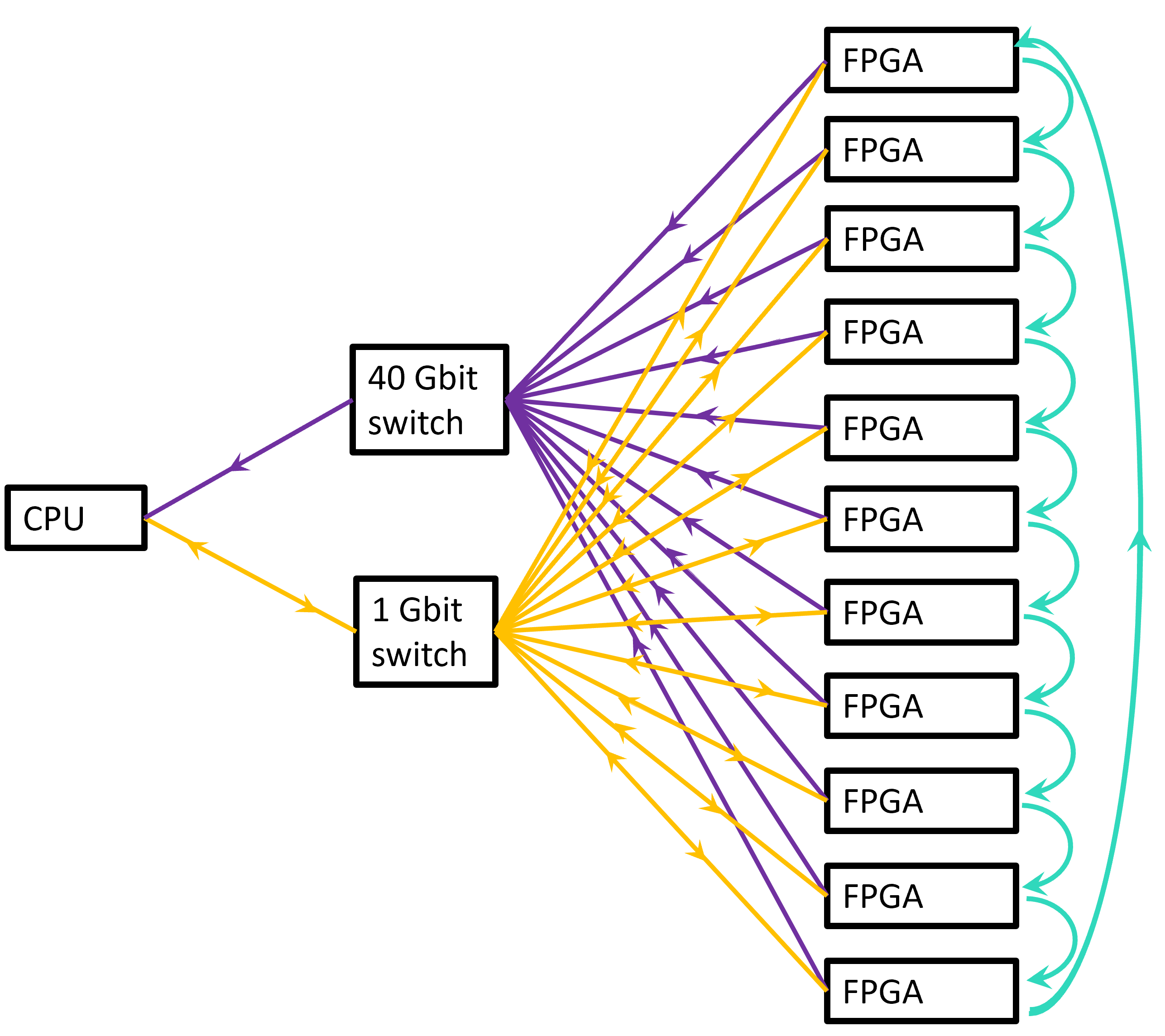}
    \caption{Flowchart summarizing the information flow in the cosmic ray detection system. Configuration settings and system monitoring information travel between the FGPAs and a central computer on a 1\,Gb Ethernet network (gold arrows). Timeseries data travel from the FGPAs to the central computer on a 40\,Gb Ethernet network (purple arrows). Trigger signals travel from each FPGA to its neighbor in a loop on a dedicated wire between general purpose input output pins on the FPGA boards (green arrows).}
    \label{info-flow}
\end{figure}

If the trigger condition is met for any SNAP2 board, and not vetoed by that board's RFI veto system, all the SNAP2 boards read out their buffers in order to save a snapshot for the entire array, including the antennas for which the signal remained below threshold. This will allow sub-threshold pulses to be used in the air shower reconstruction.  To accomplish this readout from the entire array, all the 11 SNAP2 boards are wired to each other in a loop linking their General Purpose Input-Output Pins, and this loop communicates a one-bit signal triggering data readout. This one-bit signal transmission is faster than Ethernet and has a fixed, predictable latency.
Figure~\ref{info-flow} summarizes the information networks in the cosmic ray detection system.

When a trigger signal halts writing to the buffer and initiates readout, the buffered 20\,$\mu$s timeseries for each individual-dipole signal chain is read out into UDP Ethernet packets and transmitted from a QSFP+ Ethernet port to a dedicated computer via the same 40\,Gb Ethernet switch that also serves the astronomy correlator.  After a 20\,$\mu$s snapshot is read out over Ethernet, the FPGA resumes writing to the buffer and awaits new triggers. The readout deadtime is set by how fast the receiving computer can process the Ethernet packets, rather than the maximum speed that the SNAP2 boards can transmit the data. Every readout causes a detector dead time of 0.7\,ms. Readout rates and dead time are discussed in more detail in section \ref{sec:detectorsettings}.

The cosmic ray system's main use of FPGA resources is BRAM for the voltage buffer and DSPs for the FIR filter.  Since much of the trigger logic operates on Boolean signals, resource use is small in this part of the system. The firmware was developed and compiled using tools from the CASPER collaboration \cite{Hickish2016Casper}.  For more detailed documentation on the firmware design, see \cite{mythesis}.

\subsection{On-FPGA RFI Veto}\label{sec:fpgaveto}

The proof-of-concept cosmic ray observations prior to the upgrade \cite{Monroe2020} encountered impulsive RFI at an estimated rate of 500 Hz. This background necessitates that routine observations have a first stage of impulsive RFI event rejection in the FPGA trigger firmware, to keep the trigger rate low enough to avoid substantial Ethernet packet loss in the 40\,Gb network or at the receiving computer, and to minimize readout dead time.  Later stages of RFI rejection occur in software processing on CPUs after the snapshots of data have been transmitted off the FPGAs.

\begin{figure}[tb]
    \centering
    \includegraphics[ width=0.9\textwidth]{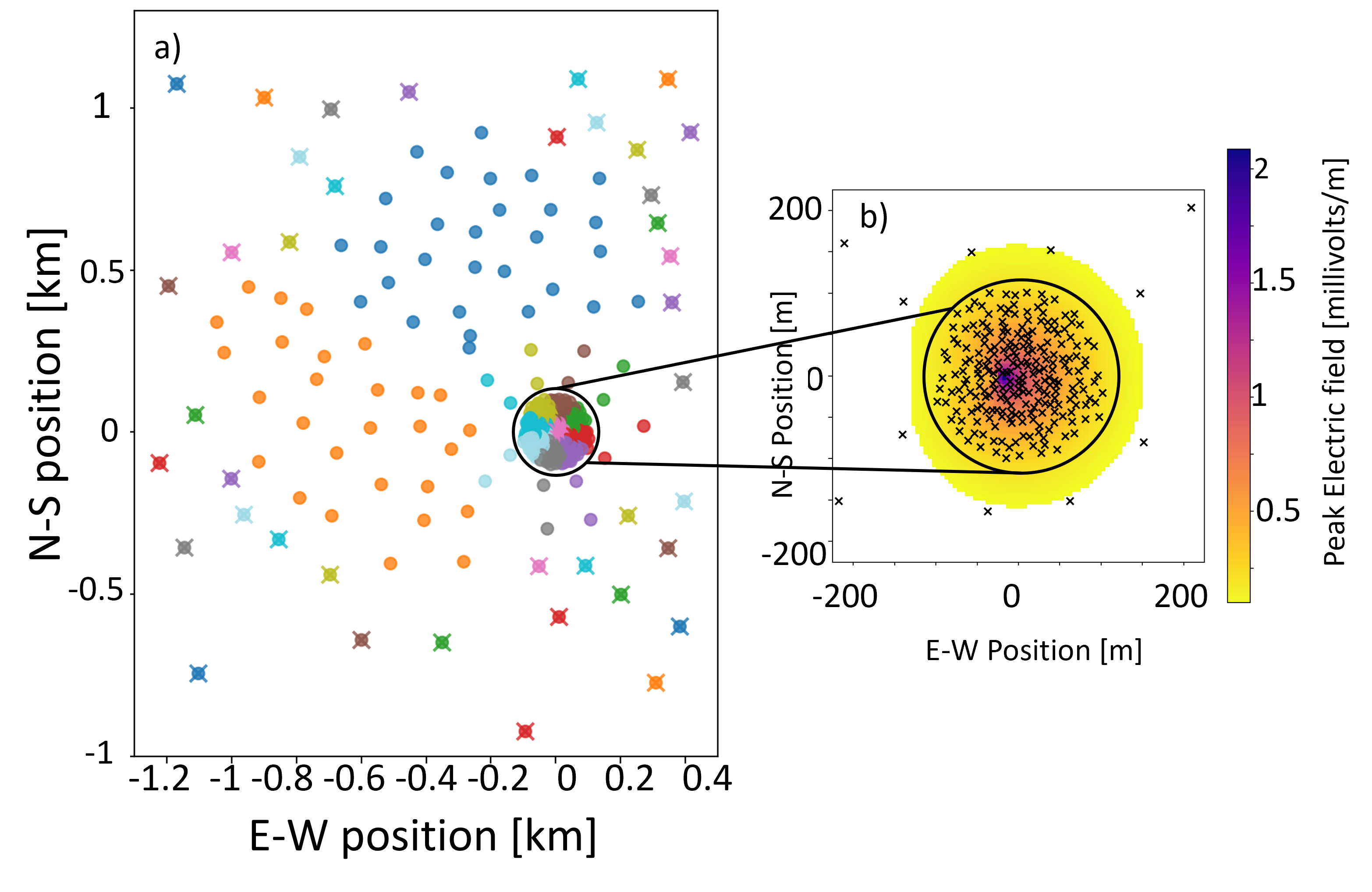}
    \caption{A) Array layout color-coded according to which FPGA processes data from each antenna.  Antennas with X markers veto RFI. B) Inset on the core array.  In this panel, to better show the densely-spaced core array all antenna locations are marked X and participate in the trigger system, not the veto. The color scale shows emission from a simulated air shower for a 100 PeV cosmic ray arriving from zenith.  }
    \label{array-layout}
\end{figure}

 In the FPGA firmware, the digital FIR filter rejects persistent RFI below 30 MHz and in the FM band, but broadband, impulsive RFI requires a different strategy. Since cosmic ray air showers are beamed, whereas many RFI sources illuminate the full array, distant antennas can veto RFI within the firmware (Figure \ref{array-layout}). Prior to the upgrade, in the 256-antenna LWA, antennas were grouped to FPGAs in sets of 16 arranged in North-South stripes and only the core array was used for the cosmic ray search demonstration. The upgrade to the 352-antenna array made all of the antennas available for the cosmic ray detector, and also offered the opportunity to optimize the groupings of antennas to FPGAs for cosmic ray detection, since the choice of assignment of antennas to FPGAs does not affect the other science goals of the array. Figure \ref{array-layout} shows the layout for the upgraded array, color-coded according to which of the 11 SNAP2 FPGA boards processes data from each antenna. For antennas in the core array,  groups of neighboring antennas are processed by the same FPGA, dividing the core into eight wedges and a central region (see Figure \ref{array-layout}).  This arrangement increases the probability that an air shower will illuminate enough antennas within the same FPGA to meet the trigger condition.  Most of the sparse array is split into two large regions and assigned to the remaining two FPGAs.

Each of the eleven FPGAs also processes a few of the most distant antennas, separated by a sufficient distance that most cosmic ray air showers in the energy range of interest would not illuminate the distant antennas on both sides of the array.  Within each FPGA, when the trigger condition is met in the core antennas, the trigger is canceled if an above-threshold signal is also detected by a threshold number of the distant veto antennas, within a suitably larger time window to account for the light travel time across the full array. If the trigger is not vetoed, then the buffered snapshot of data is transmitted off the FPGAs.  Importantly, the veto system requires that multiple veto antennas from the same FPGA detect an event in order to reject it, thus requiring detections by veto antennas on opposite sides of the core. Note in Figure \ref{array-layout} that some of the antennas marked as veto antennas to the south of the core array are in fact near enough to the core to be illuminated by some air showers that illuminate the core. The FPGAs that use these veto antennas have at least one other veto antenna that is more distant on the opposite side. The antennas on the north side are far enough that only extremely inclined air showers or air showers from cosmic rays with energies well above the second knee would simultaneously illuminate these veto antennas and the veto antennas south of the core.
The performance of the veto system is discussed in the next section, and shown in figure \ref{thresholds}.

\subsection{Optimizing detector settings }\label{sec:detectorsettings}

Operation of the cosmic ray detector is controlled via a software interface that sends commands to read and write certain dedicated registers in the firmware.  This software interface can also read various rate monitoring statistics directly from the SNAP2s.  Configuring the cosmic ray detection system includes setting: network address information, the assignment of specific antennas to trigger or veto roles, trigger and veto window lengths, power thresholds for trigger and veto antennas, and the coincidence thresholds for the trigger or veto.  The assignment of the antenna roles and the trigger and veto window lengths are fixed according to the array geometry. 

Routine operation of the cosmic ray detector requires threshold settings that maintain a manageable data rate while also operating with acceptable sensitivity to cosmic rays.  The threshold settings are chosen to satisfy requirements for detector stability and dead time. There are four thresholds in question: the power threshold for core antennas, the power threshold for veto antennas, the number of core antennas that must be above threshold to cause a trigger, and the number of veto antennas that must be above threshold to cause a veto. The two requirements for routine operations are an average readout rate of fewer than 50 events per second, and that the dead time resulting from time that the veto antennas are blocking RFI be less than 10\%.  Both of these requirements apply to average conditions over several hours. The data network can handle isolated bursts of data at rates of at least 180\,Hz.

Settings that satisfy these criteria were identified by iterating through a series of combinations of settings and then measuring event rate statistics logged by the firmware, and repeating over the course of a night. These statistics include the core trigger rate, veto rate, readout rate, veto dead time, etc.  These measurements are made by reading the rates directly from the SNAP2, without saving the Ethernet packets of time series data output by the cosmic ray detector and thereby allowing rate measurements that explore a parameter space that extends into rates that would overwhelm the data network. The SNAP2 boards also log the rate that each antenna exceeds the trigger and veto power thresholds. 

\begin{figure}[tb]
    \centering
    \includegraphics[ width=0.9\textwidth]{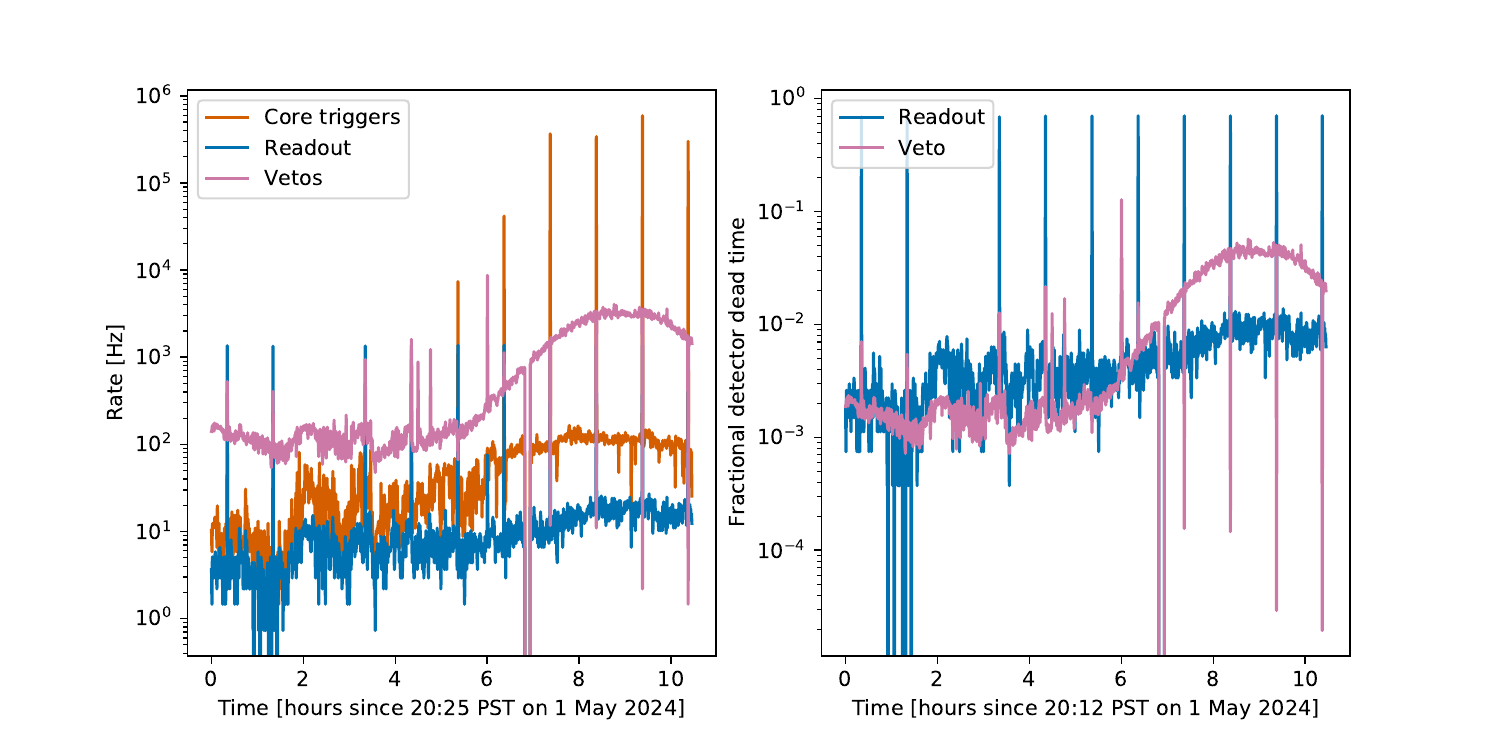}
    \caption{ Left: Rates of triggers, vetos, and readouts over the night of 2024 May~1--2 , using the threshold settings that were identified to meet the requirements for veto dead time and readout dead time. These rates are measured individually per FPGA board and the median across all the FPGAs is shown. The peak in the smooth component of the veto curve corresponds to the time that the thermal noise from Galactic synchrotron emission is highest. Right: Dead time fraction due to the veto and the readout, over the same night, with the same threshold settings. Again, the median across all FPGAs is shown. } 
    \label{thresholds}
\end{figure}

The threshold-coincidence trigger pushes the thermal limit of the system much lower than the individual-dipole threshold-excess rates.  The probability of a thermal coincidence decreases as the power of the number of dipoles required in the coincidence, which is higher for the triggers than the vetos (since there are only a few veto antennas per SNAP2 board). For the settings that meet the dead time criteria, the rate of coincidences among the trigger antennas is RFI-dominated and the veto rate is thermally dominated outside the times of intense RFI.  Figure \ref{thresholds} shows that the veto rate peaks at the time that the noise from Galactic synchrotron emission is highest.  The veto system lowers the trigger rate by an order of magnitude compared to the raw trigger rate (Figure \ref{thresholds} left panel).   The veto and readout deadtime shown in the right panel of Figure \ref{thresholds} shows that these threshold settings meet the requirements on the dead time.  The veto and readout each contribute a few percent average dead time.

Strong daytime RFI requires additional analog attenuation to avoid saturating the digitizers.  In the near future, we plan to commission cosmic ray detector threshold settings that compensate for this stronger analog attenuation to allow daytime cosmic ray searches. 

\subsection{Event Classification in Software}\label{sec:classify}

\begin{figure}
    \centering
    \includegraphics[ width=0.3\textwidth]{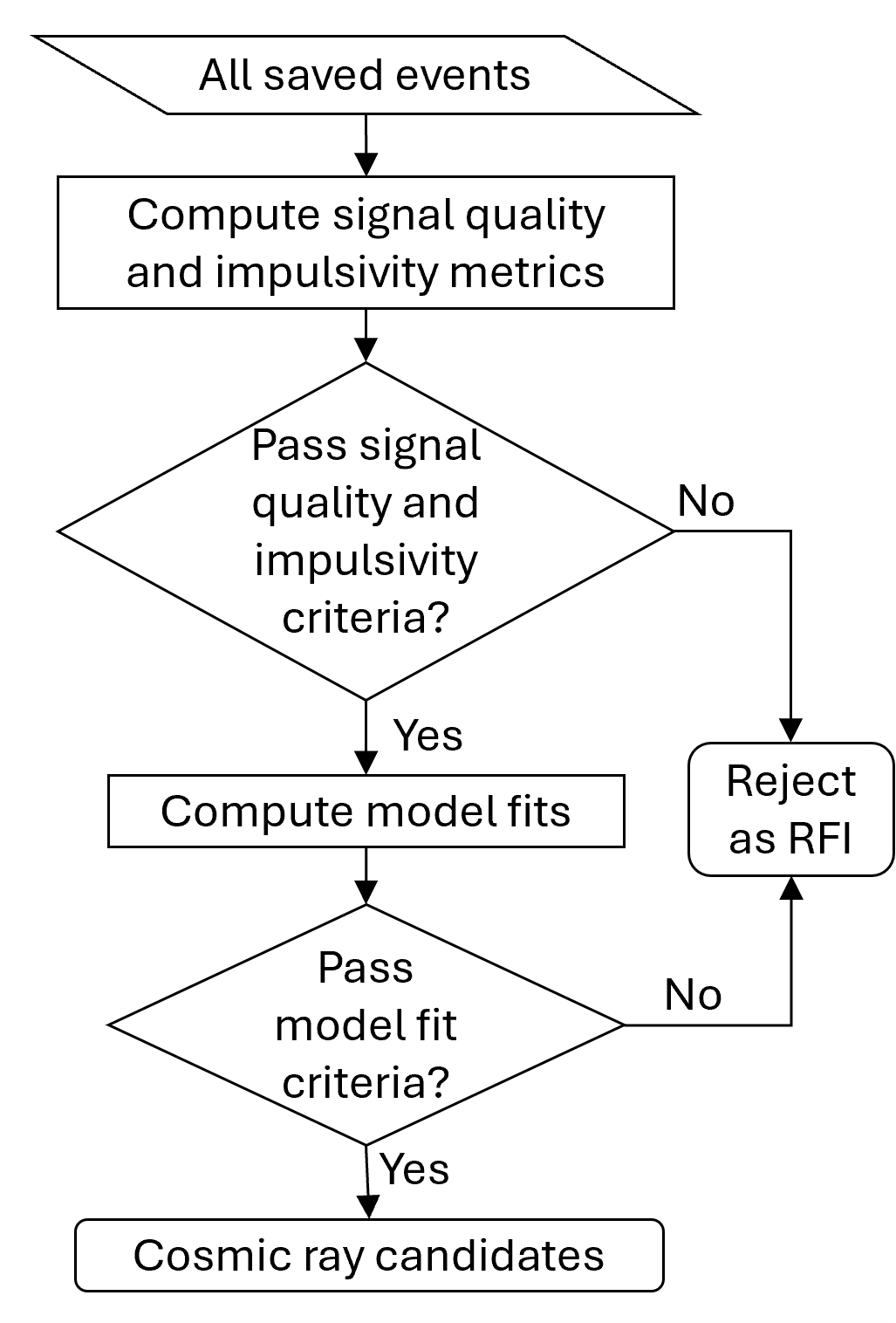}
    \caption{ High-level flow chart of the software processing for RFI-rejection and cosmic ray candidate identification. } 
    \label{fig:flow}
\end{figure}

Not all RFI is rejected by the veto system, because some RFI does not illuminate the full array, and because stochastic noise fluctuations mean that RFI with field strengths near the detection threshold will be above threshold for some antennas and below threshold for others. The impulsive RFI in the environment of the OVRO-LWA consist of sources on the horizon (particularly arcing power transmission lines), sources on the ground nearer to the array, and sources in the sky. The latter are largely reflections and transmissions from airplanes. Events for which waveform data are transmitted off the FPGAs undergo classification as RFI or cosmic ray candidates using a software pipeline, written in Python, which has two main parts: basic signal quality cuts, and cuts based on simple model fits (Figure~\ref{fig:flow}).

The first set of cuts rejects events with major signal quality problems that can be caused by intense \hbox{RFI}, and rejects events for which the waveforms are not impulsive. For each event, a power time series is computed for each antenna by applying a bandpass FIR (the same as that used in the trigger firmware) to the raw voltage time series and squaring the output. Then the mean power, excess kurtosis of the filtered voltage time series, and the number of saturated samples (samples that reach the limit of the ADC output, thereby clipping the true signal) are computed. 

The quality criteria applied to each of the timeseries from each individual-dipole signal chain require that the timeseries must have fewer than ten saturated samples, an excess kurtosis between~$-1$ and~$1$, and a mean power between 225 and 2500 ADC units squared.  The saturation limit is set at 10 samples because bright air showers can saturate antennas near the shower core for a few samples.  The kurtosis criterion identifies signals that significantly deviate from Gaussian-distributed stochastic noise in the period before the event. The power limits reflect the typical range of powers for properly functioning signal chains in the absence of RFI, considering the differing attenuation in cables of different lengths.  High power outside the acceptable range can indicate a long-duration RFI event, and low power can indicate a fault in the signal chain.  

At any given time, some signal chains are in need of maintenance, and so the quality criteria applied to each event do not require every individual signal chain to pass the signal quality criteria.  
In order to pass the basic signal quality cuts, fewer than 10 dipoles' timeseries can exceed the limit on saturated samples. The event must also have fewer than 10 dipoles for which the excess kurtosis lies outside the range $-1$ to~1, and fewer than 200 dipoles can have mean powers outside the range 225 to~2500 ADC units squared.  The choice of 10 dipoles for the kurtosis and saturation cuts reflects an upper bound on the typical number of dipoles with faults producing these symptoms in their signal path (usually these are only a few signal chains).  The tolerance of a larger number of dipoles ($<$200) outside the normal power range allows the search to operate during times when the outer portion of the array is powered off.  At the end of the commissioning process, these outages have become very rare.  For events that pass these basic signal quality cuts, the individual faulty signal chains that fail the cuts are excluded from subsequent analysis.

As a computationally simple discriminant between largely impulsive events and extended waveforms, the ratio of the power before and after the waveform peak is computed. The peak is identified as the time sample where the maximum of the absolute value of the analytic signal computed from the filtered voltage timeseries occurs. The databuffer on the SNAP2s is long enough that the first half (first 2000 samples) sample the background before the event. The ratio of power before and after the peak, $P_r = \frac{P_1}{P_2} $, is computed where $P_1$ is the mean of the first 2000 samples in the power timeseries (square of the filtered voltages) and $P_2$ is the mean of the 50 consecutive samples beginning 25 samples after the peak of the analytic time series. The offset of 25 is chosen to exceed the impulse response time of the analog and digital filters. 
For a waveform consisting of an impulse overlaid on Gaussian noise, the expectation value of this power ratio is 1. 
The next step computes the median of $P_r$ for all the signal chains with a peak signal to noise ratio  (S/N) greater than 6, after removing signal chains that fail the quality criteria discussed above.  This median is computed separately for the two polarizations.  To pass the impulsivity cut, the median $P_r$ must lie within 0.8 and 1.1 for BOTH polarizations.  This impulsivity cut could in principle be moved to an implementation in the firmware trigger in the future.

All events that pass the above signal quality cuts proceed to the second part of the classification, which involves model fits to the peak S/N and the time of this peak at each antenna. 
A model of a spherical wavefront is fit to the times of the peak at each antenna's position in order to identify nearby RFI sources. This model approaches a plane wave for distant RFI and cosmic rays, and provides a direction estimate but not a distance estimate in these cases. Prior to the fit, the recorded time of the peak electric field is corrected to compensate for delays in the signal path unique to each antenna.  Section~\ref{sec:cal.delay} discusses the details of this delay calibration.  In order to prevent the possibility of delay calibration errors in a few antennas influencing the event direction measurement, outliers are iteratively identified and excluded for several iterations of least-squares fitting. After computing residuals for a best-fit wavefront model, the subsequent fit iteration excludes those antennas for which the residuals deviate from the median residual by more than four times the median absolute deviation. The threshold of four times the median absolute deviation separates antennas that have a high residual due to delay calibration error with a low probability of cutting off the tails of the residual distribution for antennas with correct delays.   Events that cannot be well described by an arriving wavefront from a well-defined direction are rejected.  The criteria for a reliable direction reconstruction are that the fit routine completed, the best-fit model has an RMS arrival time residual of less than two clock cycles, and more than 15 dipoles (of the dominant polarization only) remain un-flagged after the final iteration of the fit.

The second model fit searches for events with a spatially concentrated power distribution. After exploring several approaches for a computationally simple metric for the spatial concentration of air shower radio emission, we adopt an approach of fitting a 2D elliptical Gaussian function to the measured S/N as a function of antenna position. Again, this is a very simple approximation of an air shower lateral distribution function, aimed only to distinguish the beaming of cosmic ray air showers vs.\ RFI events.

The metric used throughout for the S/N is the maximum magnitude of the analytic signal (Hilbert envelope) divided by the RMS from the first 2000 samples.  For this S/N metric, the typical S/N of pure noise is 3.8. Both the spatial distribution fit and the arrival direction fit only use antennas with a peak S/N greater than 5.5. The fit is performed using the S/N values for the stronger of the two polarizations, defined as whichever polarization has a larger mean S/N for the event, among signals with an S/N above 5.5.

 The fit to the spatial distribution of the peak S/N returns the optimized values of the overall amplitude (A), the position of the shower core ($x_0$ and $y_0$, defined in meters relative to a reference position in the array), the orientation of semi-major axis of the elliptical contours ($\phi$), the scale parameter $\sigma_x$, and the aspect ratio $\frac{\sigma_y}{\sigma_x}$ for the following Gaussian approximation to a lateral distribution function:

\begin{subequations}\label{eq:y}
\begin{align}
\label{eq:y:1}
f & =A e^{-a(x-x_0)^2 +2b(y-y_0)(x-x_0)+c(y-y_0)^2}   
\\
\label{eq:y:2}
a & = \frac{\rm{cos}^2\phi}{2\sigma_x^2} + \frac{\rm{sin}^2\phi}{2\sigma_y^2}
\\
\label{eq:y:3}
b & = -\frac{\rm{sin}\phi\rm{cos}\phi}{2\sigma_x^2} + \frac{\rm{sin}\phi\rm{cos}\phi}{2 \sigma_y ^2}
\\
\label{eq:y:4}
c &= \frac{\rm{sin}^2\phi}{2\sigma_x^2}+ \frac{\rm{cos}^2\phi}{2\sigma_y^2}
\end{align}
\end{subequations}

To avoid multiple equivalent solutions, the parameter space is restricted to $\frac{\sigma_y}{\sigma_x} \geq 1$ and $ 0 \degree \leq \phi \leq 180 \degree$, such that $\sigma_x$ is always the scale parameter along the semi-minor axis and increasing $\phi$ rotates the semimajor axis East of North (equivalently West of South).
Event files are processed in parallel using Gnu Parallel \cite{Tange2018gnu}.

  \begin{table}
      \centering
      \begin{tabular}{|c|c|c|} \hline

        Sample Cut & Events & Fraction \\ \hline
         Total & 3828175 & 1 \\ 
         Pass signal quality cut & 2839284 & 0.74 \\ 
         Pass impulsivity cut & 105270 & 0.027\\ 
         Pass quality \& impulsivity & 92308 & 0.024 \\ \hline
         Wavefront direction fit succeeds & 86343 & 0.023 \\ 
         Both fits succeed & 1767 & 4.6 x $10^{-4}$ \\ 
         Fits succeed and lateral scale \textless 500m & 15 & 5.28 x$10^{-6}$  \\ 
         Pass all model fits cuts & 8 & 2.09 x$10^{-6}$  \\ 
        \hline
    \end{tabular}
      \caption{Selection criteria applied to a dataset collected from 48.7 hours of commissioning data. The solid line separates the analysis that is performed on all events from the analysis performed only on the events that pass the signal quality and impulsivity cuts, but the fraction column continues to be the fraction of the original total.}
      \label{tab:cuts}
  \end{table}

Based on the results of the model fits, the following cuts are applied to identify candidate cosmic rays.  The first cut simply checks that both fit routines completed successfully, and that the Gaussian fit found a lateral scale larger than 50 m (to help exclude RFI that originates within the array and illuminates only a small region of antennas). For most RFI events, the Gaussian fit does not converge, or converges to extreme parameter values when it attempts to fit a 2D Gaussian to an inherently flat distribution.  A cut is applied on how well the Gaussian model describes the data, by requiring that the RMS residuals be less than 2 (in units of S/N, which for thermal noise have an uncertainty of 1).     The remaining events have a well-defined wavefront-fit arrival direction and parameters describing their lateral distribution function as a 2D Gaussian approximation. Events within 15 degrees of the horizon are discarded, because uncertainties in the antenna beam model complicate the accurate reconstruction of air showers near the horizon, and because RFI from nearby ground-based sources can create localized emission patterns that pass the Gaussian fit criteria.

  \begin{figure}[tb] 
    \centering
    \includegraphics[ width=0.6\textwidth,trim={0cm 0cm 0cm 1cm},clip]{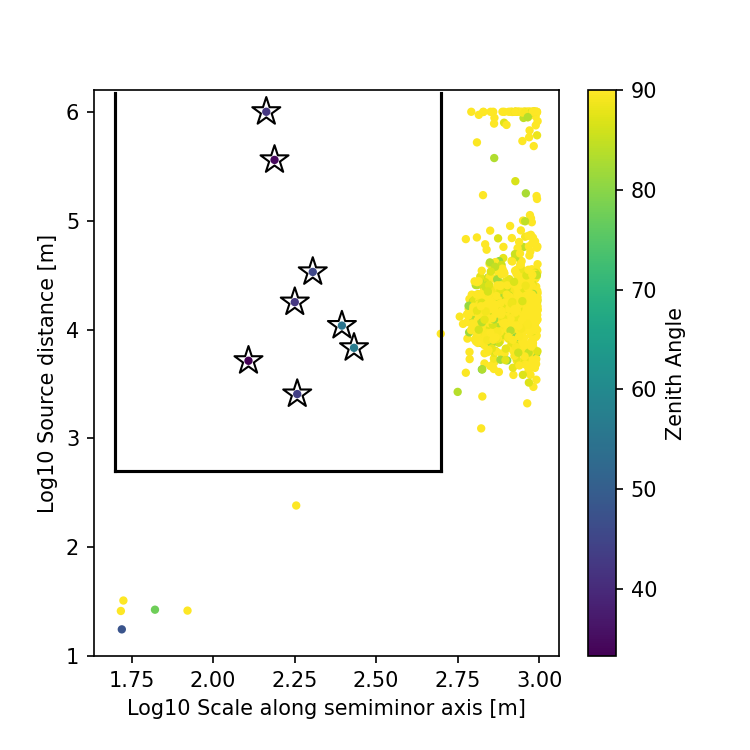}\vspace{-0.5cm}
    \caption{ Events passing the criteria for successful model fits to the arrival times and spatial illumination pattern are shown in a parameter space of Log10 lateral scale (abscissa), Log10 source distance (ordinate axis), and zenith angle (color).  Black lines mark the selection cuts on the source distance and the size scale of the best-fit Gaussian. Events that additionally pass the zenith angle cut (zenith angle $<75\degree$) are highlighted with black stars and are the cosmic ray candidates.  } 
    \label{fig:cuts}
\end{figure}

The lateral scale of the best-fit Gaussian must be smaller than 500\,m, again excluding RFI with a roughly flat distribution or local non-Gaussian gradient.  The source distance must be larger than 500\,m to exclude near-field RFI originating within the array. 

Table~\ref{tab:cuts} summarizes the cosmic ray candidate selection process and shows the number of events, as well as the fraction of the total number of saved events, that pass each criterion, for a dataset of 48.7~hours of observations collected at night time during six nights between April and October 2024.

\begin{figure}[tb]
    \centering
    \includegraphics[ width=0.9\textwidth]{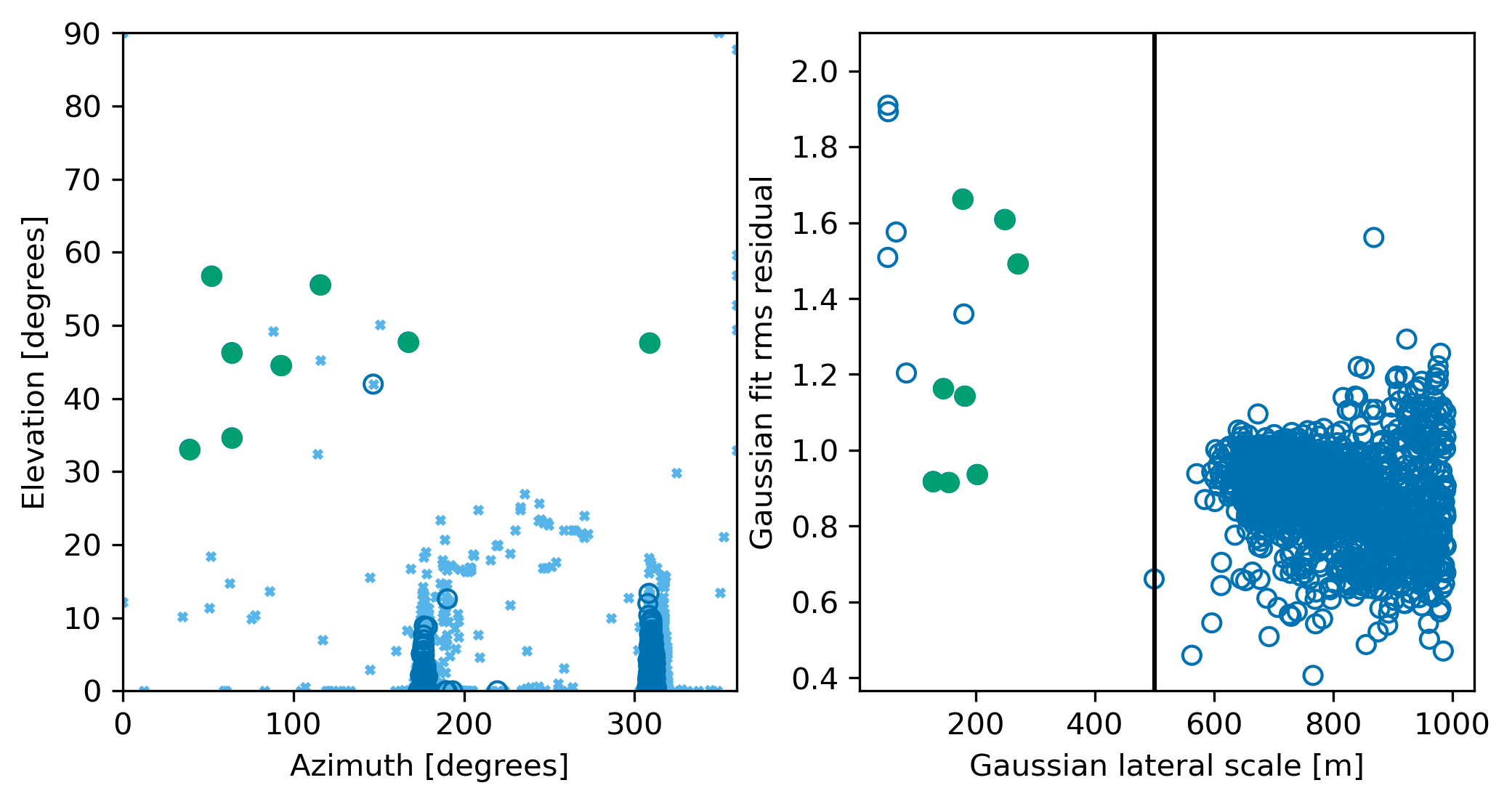}
    \caption{Impulsive events in the 48.7 hour dataset. Green filled circles mark the events that pass the cuts identifying cosmic ray candidates. Open blue circles mark RFI events ultimately failing the selection criteria but for which the Gaussian fit succeeded.  \textit{Left:} Azimuth and elevation of the cosmic ray candidates and RFI events. Light blue X's mark all the RFI events for which the wavefront fit succeeded but the Gaussian fit did not. \textit{Right:} RMS residual of the Gaussian fit and the lateral scale of the best-fit Gaussian, for events passing the basic quality criteria on the Gaussian fit. The solid black line marks the selection cut for the lateral scale.
    } 
    \label{selection}
\end{figure}

Eight cosmic ray candidates are identified from this dataset by these cuts. Figures \ref{fig:cuts} and ~\ref{selection} show the results of the model fits for events in this dataset.  This dataset is presented because the detector settings were constant during these observations, demonstrating the performance of the detector over a period of several months.  This dataset has an average rate of detections of one cosmic ray candidate per six hours of observing.  During the commissioning process, an additional six cosmic ray candidates were identified that pass the selection criteria presented here but were detected during times that the detector settings and array conditions were in varying states of sensitivity.  Thus, these are excluded from rate estimates but included in the total sample of fourteen  OVRO-LWA cosmic ray candidates.

Figure \ref{fig:cuts} shows the Gaussian spatial scale, source distance, and zenith angle for all of the events that have successful model fits and thus have well-defined Gaussian spatial scales and arrival directions. Black lines mark the cuts on the Gaussian spatial scale and the source distance, and black stars mark the events that additionally pass the zenith angle cut (with zenith angle shown as the color of the points).  Sources beyond a few kilometers are nearly indistinguishable from plane waves, and thus distance estimates cease to be useful beyond this scale. In the future, the distance bounds of the fit parameter space could be made much smaller than those used in the fits in Figure \ref{fig:cuts}.

 As seen in Figure~\ref{selection}, most of the RFI originates from two sources near the horizon, likely aligned with faulty power transmissions lines.  Ongoing efforts to monitor and have these power lines repaired as new faults appear is improving the RFI environment.  Monroe et al.~\cite{Monroe2020} observed, in addition to power lines, a large number of events coinciding with air plane tracks.  The on-FGPA veto may be rejecting these events before they are saved, whereas the power line RFI propagating along the ground manages to illuminate the array unevenly.  
\begin{figure}[tb]
    \centering
    \includegraphics[ width=0.9\textwidth]{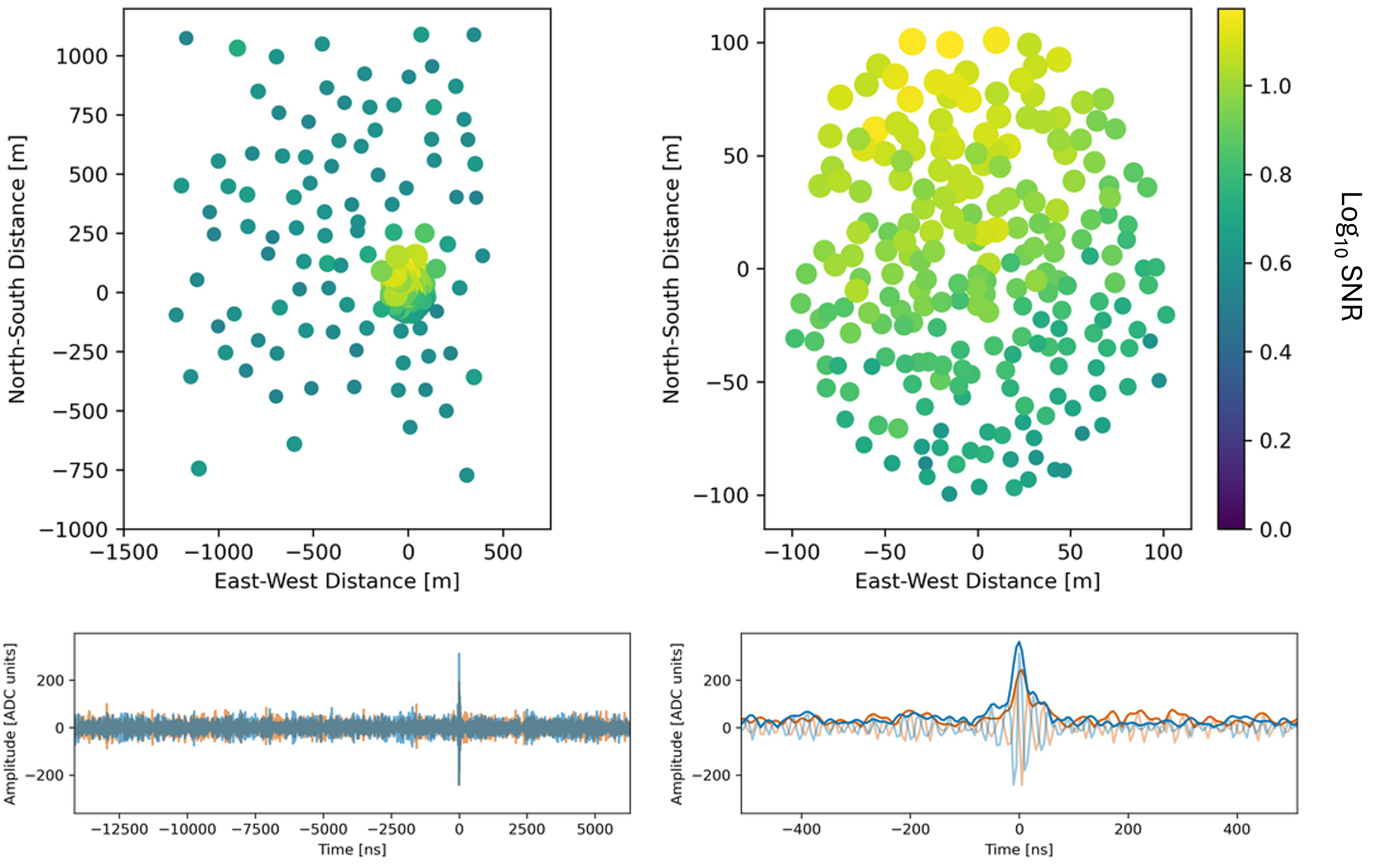}
    \caption{An air shower candidate from 2024 May 1 over the core array, with an arrival direction zenith angle $33^\circ$, and azimuth $52^\circ$.  \textit{Upper left:} Log signal-to-noise ratio (for the dominant polarization) at each antenna location, showing the beamed emission pattern of the air shower candidate. \textit{Upper right:} Same as upper left but zoomed in on the core array. \textit{Lower left:} Waveforms from both polarizations of the antenna with the brightest signal. \textit{Lower Right:} Same as lower left but in a narrower time range around the peak.  The orange and blue represent the two polarizations, with the Hilbert envelope in bold.} 
    \label{may1}
\end{figure}

\begin{figure}[tb]
    \centering
    \includegraphics[width=0.95\textwidth]{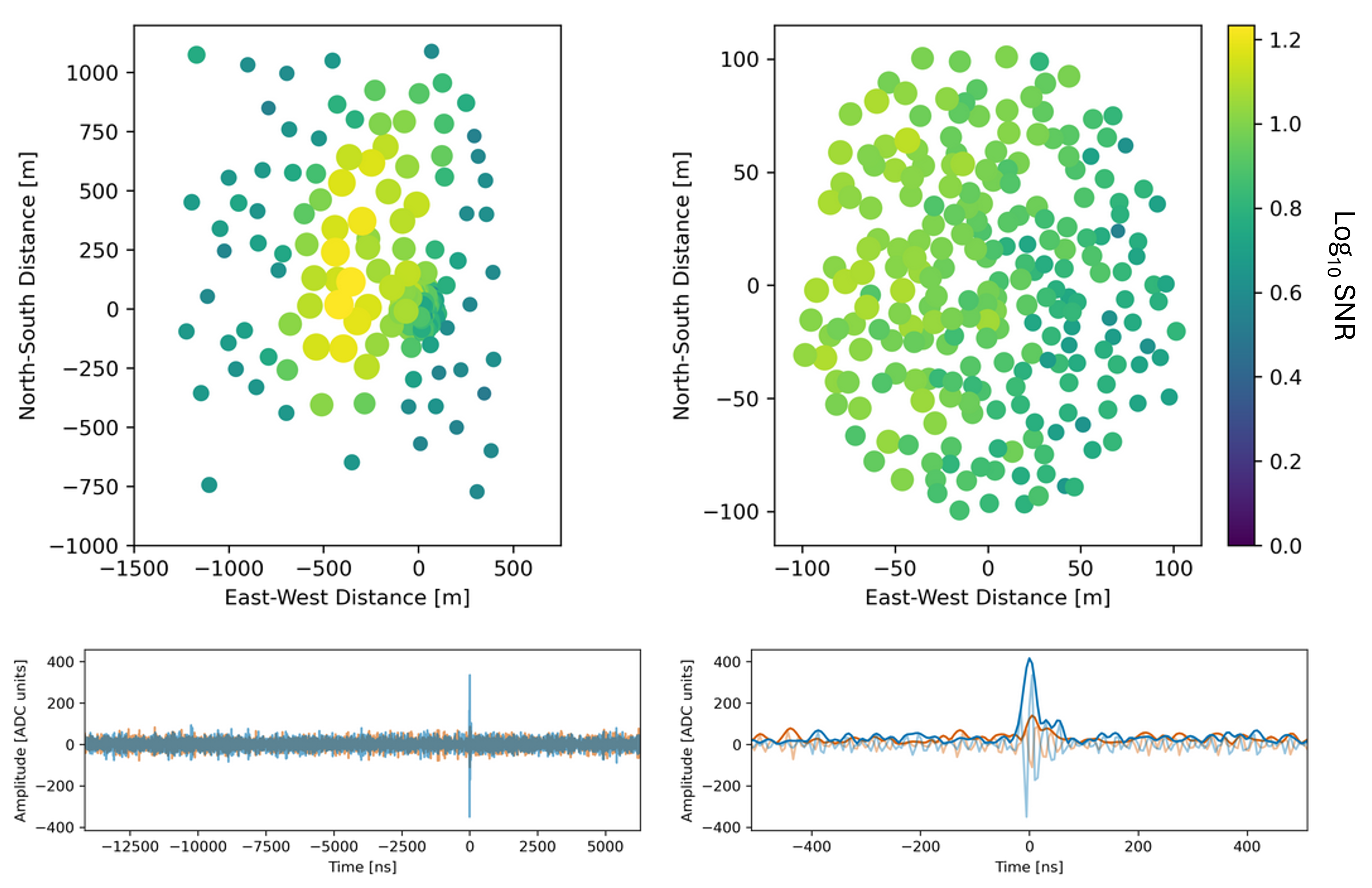}
    \caption{Example of a more inclined air shower candidate from 2024 February~21 illuminating a larger part of the array, with an arrival direction zenith angle 63\arcdeg\ and azimuth 190\arcdeg. See Figure~\ref{may1} for a description of each panel.} 
    \label{feb21}
\end{figure}

Figures~\ref{may1} and~\ref{feb21} show examples of two of the cosmic ray candidates.  The first example (Figure \ref{may1}) illuminated the core array and has an arrival direction zenith angle of 33 degrees.  The second example is more inclined (and likely higher energy), with a zenith angle of 63 degrees and illuminates a large region of the sparse array.  The array layout of sparse array and dense core array allows air showers in these different scenarios to both have well sampled emission footprints, which will be important for future shower reconstruction work.

\subsection{Polarization as a validation of the cosmic ray candidate sample}\label{validation}

\begin{figure}[tb]
    \centering
    \includegraphics[ width=0.6\textwidth,trim={0cm 0cm 0cm 0cm},clip]{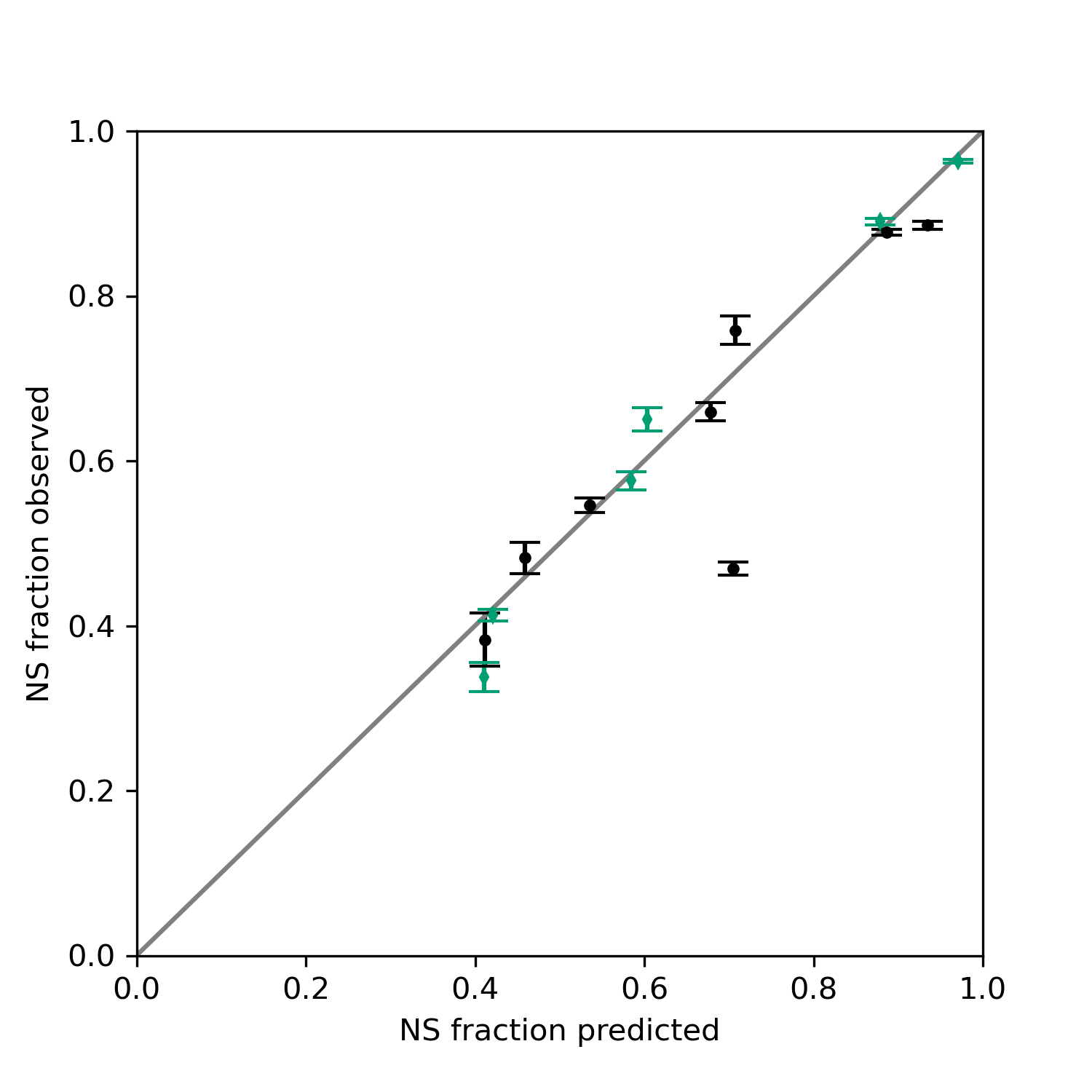}\vspace{-0.5cm}
    \caption{Each cosmic ray candidate's observed polarization compared to that predicted for air shower geomagnetic emission. Black circles and green diamonds denote the cosmic ray candidates detected in the 48.7~hour dataset (\S\protect\ref{sec:classify}) and at other times during the commissioning process, respectively. The latter category pass the same selection criteria but are excluded from the rate analysis due to non-uniform detector settings. The gray line marks observation = prediction.  The errorbars are the standard error on the mean polarization fraction among the antennas.}
    \label{pol}
\end{figure}

Since polarization information is not used as a selection cut, it provides a useful validation of the sample.  Figure \ref{pol} compares the observed polarization to the predicted polarization of geomagnetic emission from a cosmic ray air shower arriving from the direction of each of the cosmic ray candidates. Geomagnetic emission in cosmic ray air showers is stronger than the Askaryan component, unless the cosmic ray arrival direction is closely aligned with the local magnetic field, and so we use the polarization of the geomagnetic component as a rough prediction of the polarization direction of cosmic ray air showers.    The geomagnetic emission is completely linearly polarized in the $\textbf{v} \times \textbf{B}$ direction, where $\textbf{v}$ is the velocity of the primary cosmic ray and $\textbf{B}$ is Earth's magnetic field.

The Earth's magnetic field at Owens Valley Radio observatory is 22,347.1\,nT North, 4,773.1\,nT East, and 42,034.1\,nT down, obtained from a service of the National Oceanographic and Atmospheric Administration \footnote{\url{https://www.ngdc.noaa.gov/geomag/calculators/magcalc.shtml\#igrfwmm}}. The direction of $\textbf{v}$ is estimated from the reconstructed arrival direction of each event.    This polarization direction is then projected onto beam models for the North-South and East-West dipoles.  Finally,  to obtain a predicted polarization fraction, the magnitude of the north-south dipole component is divided by the square root of the quadrature sum of the magnitudes of both components.  

To measure an overall observed polarization fraction of each event, the dominant polarization (that of the NS or EW dipole) is identified as the polarization with more dipoles detecting the event with an S/N greater than 5.5, and only antennas with a detection above an S/N of 5.5 in the dominant polarization are used in the polarization analysis.  The ratio of the Hilbert envelopes of the signals in the two polarizations is taken at the time sample for which the peak occurs in the dominant polarization.  As used in calculating the prediction, we define the polarization fraction as the value of the north-south dipole component compared to the square root of the quadrature sum of values from both dipoles.  The overall polarization fraction for the event is taken as the mean value of this ratio for all antennas that were above the S/N threshold.

In Figure~\ref{pol}, the error bars on the measurement are the statistical uncertainty, obtained as the standard error on the mean of the individual antenna polarizations. Thus, events that illuminate more antennas typically have a smaller statistical uncertainty on the mean polarization fraction. Errors in the direction reconstruction introduce a statistical error of roughly $1.5 \times 10^{-4}$ in the prediction, which is too small to show in Figure \ref{pol}, and uncertainties in the beam model result in a direction-dependent systematic error on the order of 0.01. The beam models were simulated using FEKO as part of the overall OVRO-LWA development work and are accurate to about 1\% above 15 degrees from the horizon.  

In Figure~\ref{pol}, the median difference between the prediction and observation is 0.009, comparable to the median statistical uncertainty of 0.011.  The notable outlier in Figure~\ref{pol} arrives from a direction that is closer to alignment with the local magnetic field than that of any of the other candidates, such that the relative strength of the geomagnetic component is lowest. For this direction,
$ {|\textbf{v}\times \textbf{B}|}/({|\textbf{v}||\textbf{B}|}) = 0.34$, making the neglected Askaryan component more significant than for the other cosmic ray candidates.  Future analyses will use full Monte Carlo simulations, such as \texttt{Corsika}, to model the emission in detail.

\section{Calibration}\label{cal}
\subsection{Timing Delay Calibration}\label{sec:cal.delay}

\begin{figure}[tb]
    \centering
    \includegraphics[ width=0.7\textwidth]{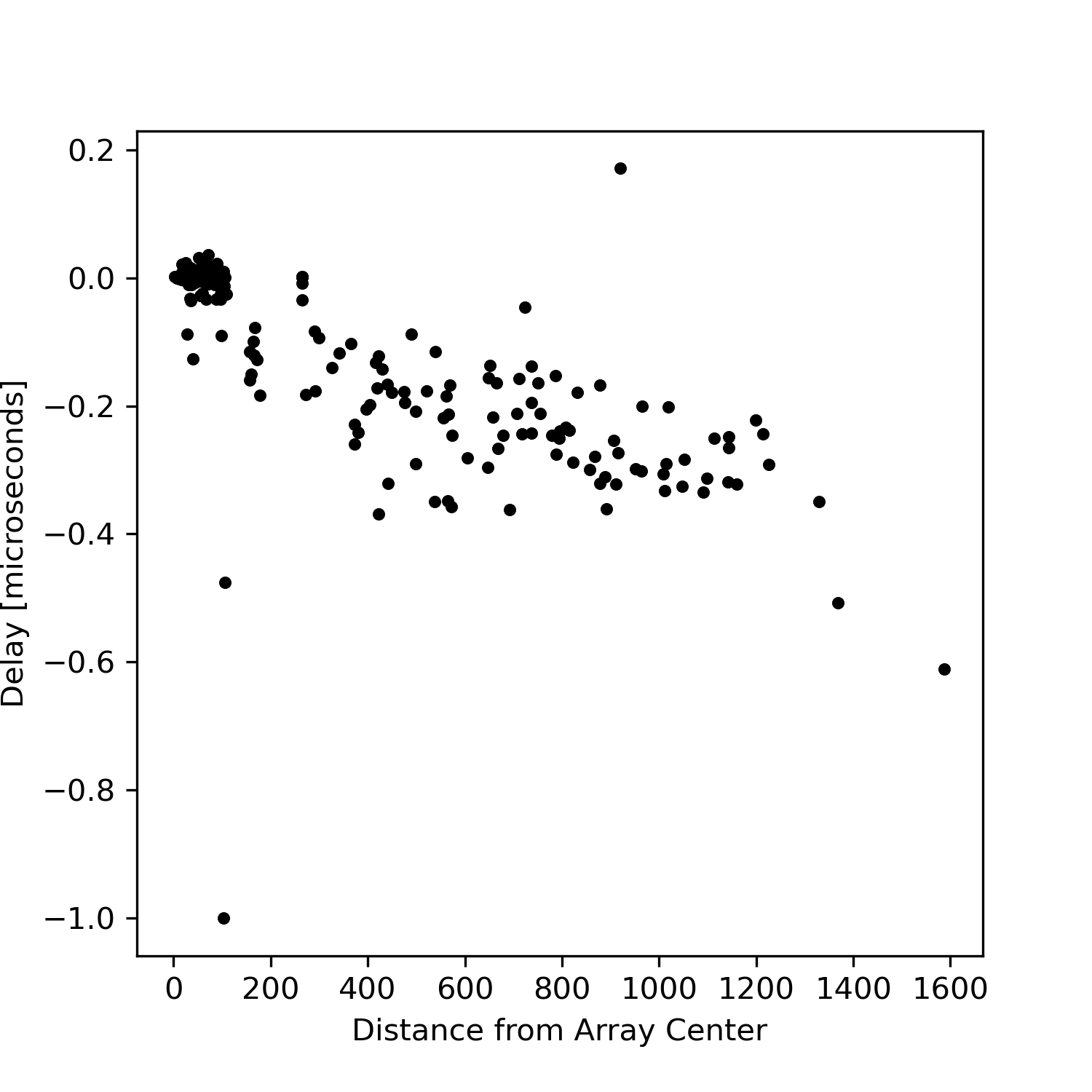}
    \caption{Delay in microseconds relative to a reference antenna in the core array, plotted as a function of antenna distance from the array center.}
    \label{delay}
\end{figure}

As described in Section~\ref{sec:coincident}, the event search firmware applies delays to the raw timeseries to adjust for the rough differences in time that signals from different antennas take to propagate through the cables and electronics to the ADCs, and these delays are estimated from the lengths and materials of the cables and fiber.
The event analysis requires a more precise knowledge of the delays between different antennas, which is calibrated using the radio astronomy imaging mode of the \hbox{OVRO-LWA}.

The same FPGAs that run the cosmic ray trigger also create a filterbank of 4096 24-kHz frequency channels using a polyphase filterbank and Fast Fourier Transform for the other observing modes and transmit 3072 of these channels, corresponding to 13.4--86.87\,MHz to a correlator implemented on GPUs.
The correlator computes the complex cross-correlation of the signals from every pair of dipoles, in 3072 of the  frequency channels.  
To calibrate the delays due to the different cables to dipoles at different antenna stands (as well as any other electronic delays), the  OVRO-LWA observes a bright radio astronomical source, such as a radio galaxy or supernova remnant, which has both a known structure and radio spectrum from which a complex cross-correlation value can be predicted. The radio galaxy Cygnus~A is a typical choice of calibrator, because it passes nearly directly over zenith. 
Comparing the predicted and observed values of the cross correlation of each antenna with one antenna selected as a reference antenna produces a complex gain coefficient for each antenna and frequency channel, relative to that reference antenna.\footnote{
The \texttt{bandpass} task within the Common Astronomy Software Applications 
\citep[\hbox{CASA},][]{McMullin2007-CASA} package is used to derive these complex gain coefficients. 
}
Any antenna free of signal quality faults may serve as the reference antenna.
The phase of this complex gain varies linearly with frequency, in proportion to the uncompensated delay between each antenna and the reference antenna.  The relative delays are solved for by finding the best-fit slope of the linear relation between the phase and the frequency.  Figure~\ref{delay} shows an example of delay solutions for each dipole.
\begin{figure}[tb]
    \centering
    \includegraphics[ width=1.0\textwidth]{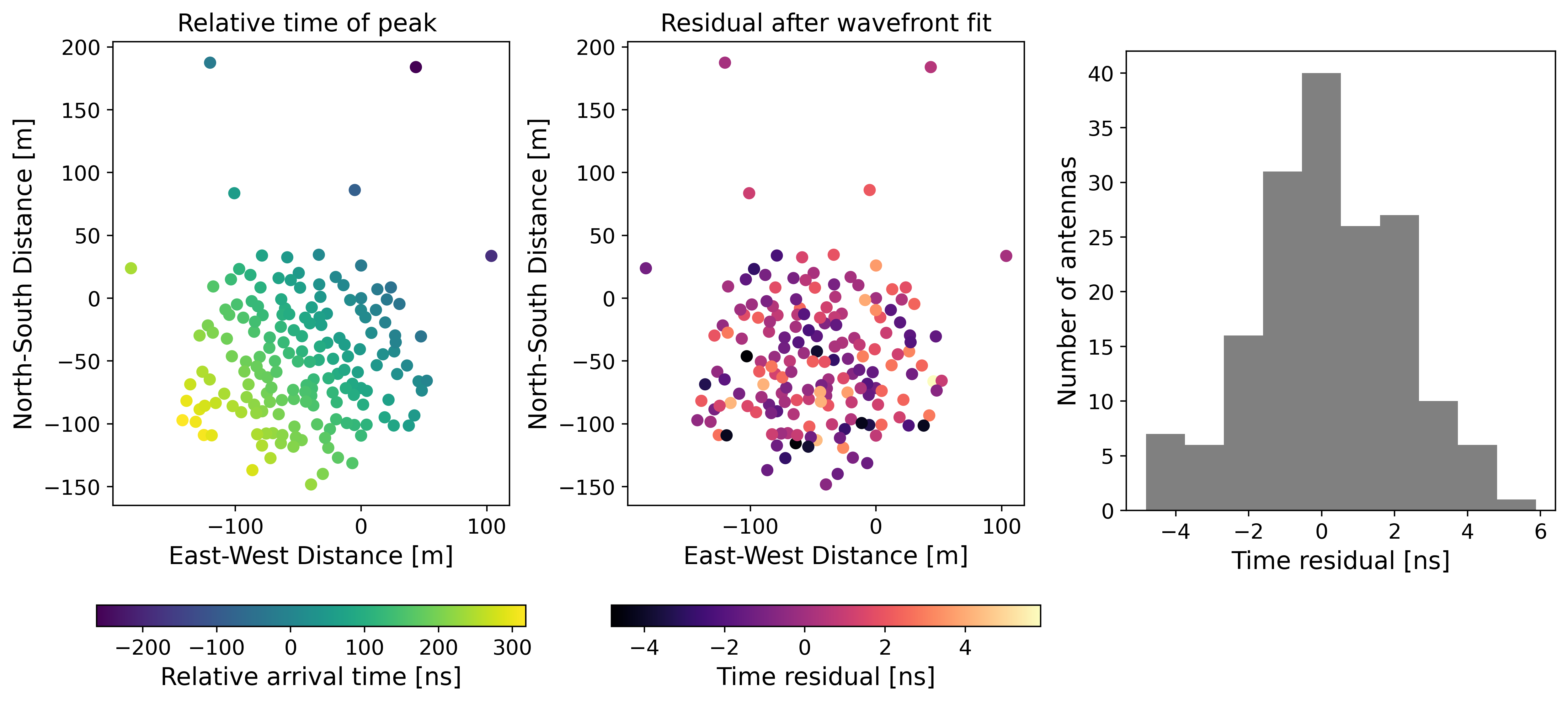}
    \caption{\textit{Left}: Wavefront arrival times for a cosmic ray air shower that illuminated the core array on May 1 2024. Delay calibration solutions have been applied. Only antennas used in the wavefront fit are shown. The spot location indicates antenna location and the color indicates arrival times. \textit{Center}: Residual difference between the measured arrival times and the times predicted by the best-fit spherical wavefront model. The best-fit arrival direction is a zenith angle of 33 degrees and an azimuth of 52 degrees. \textit{Right}: Histogram of these same residuals.}
    \label{timing1}
\end{figure}

\begin{center}
\begin{figure}[tb]
    \centering
    \includegraphics[ width=0.9\textwidth]{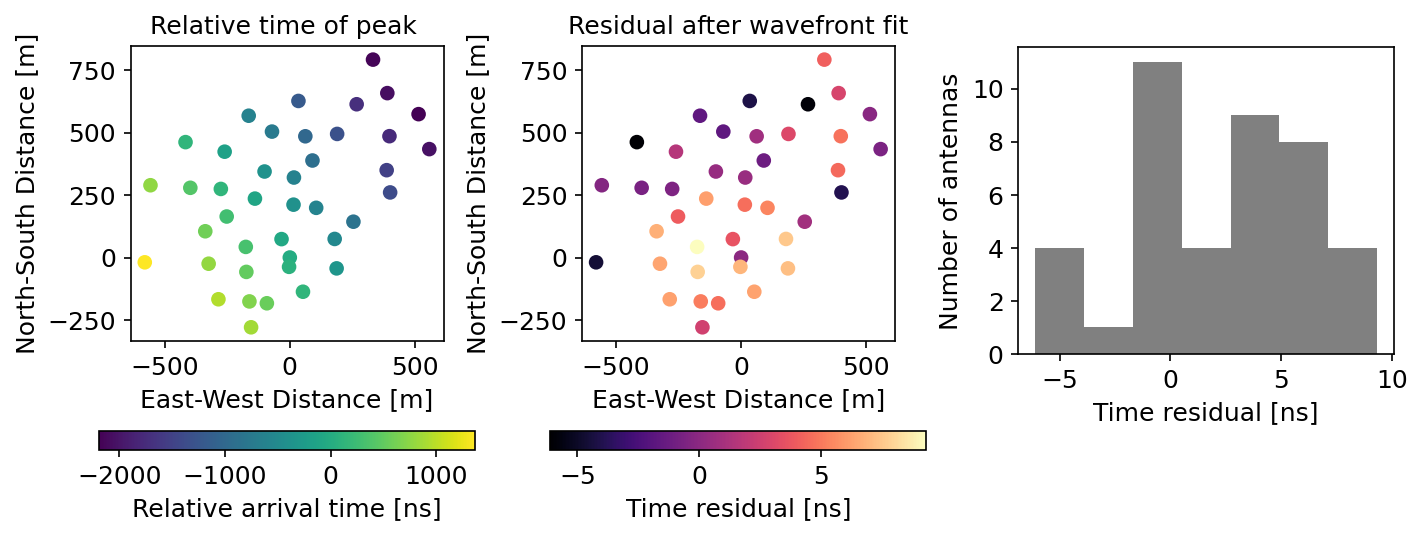}
    \caption{Left: Wavefront arrival times for a cosmic ray air shower that illuminated the sparse array on May 15 2024. Delay calibration solutions have been applied. Only antennas used in the wavefront fit are shown. The spot location indicates antenna location and the color indicates arrival times. Center: Residual difference between the measured arrival times and the times predicted by the best-fit spherical wavefront model. The best-fit arrival direction is a zenith angle of 61 degrees and an azimuth of 51 degrees. Right: Histogram of these same residuals. Larger residuals may be due to differences between the spherical wave approximation and the true, hyperbolic arrival times, which becomes more significant for air showers that illuminate a larger area of the sparse array.}
    \label{timing2}
\end{figure}
\end{center}

To calibrate the timing of events detected by the cosmic ray search, these delays obtained from the correlator are applied to adjust the time of the signal peak at each dipole. 
Figures~\ref{timing1} and~\ref{timing2} show the arrival time of the signal peak at each antenna and the residuals after subtracting the best-fit spherical wavefront model, for two observed cosmic ray air showers, one that illuminated the dense core array (Figure~\ref{timing1}) and a more inclined shower that illuminated the sparse array (Figure~\ref{timing2}).  The variance in the arrival time residuals suggest a timing precision of 2\,ns, which is less than one clock cycle of the \hbox{ADC}, and satisfies the requirement that the timing precision be better than 5\,ns. The timing residuals are smaller and more uniformly Gaussian distributed for the air shower illuminating the core array, which illuminates a larger number of antennas.

\subsection{Electric field calibration}\label{sec:efieldcal}

\begin{figure}[tb]
    \centering
    \includegraphics[width=0.9\textwidth]{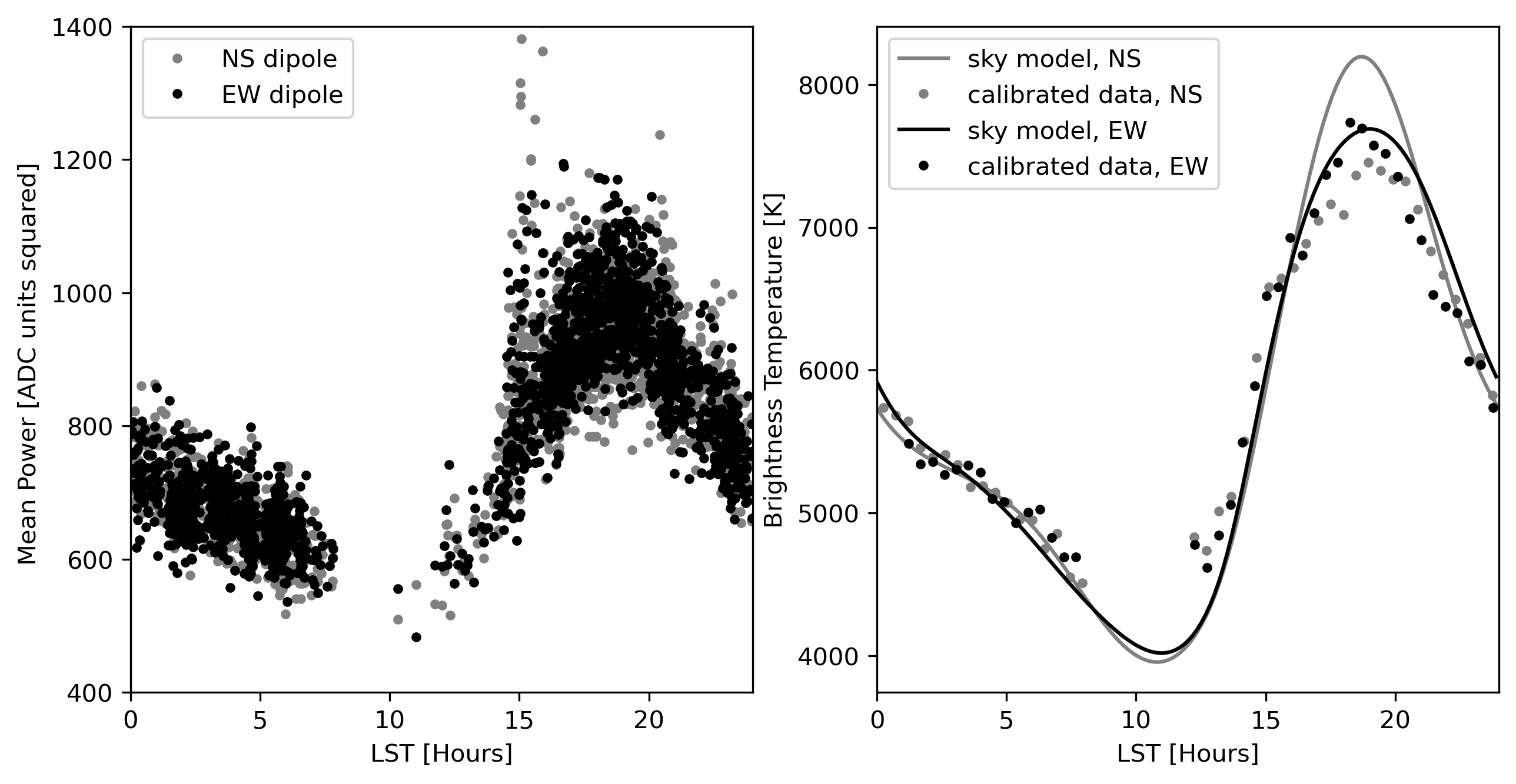}
    \caption{ \textit{Left:} Mean power (mean square of the filtered voltage timeseries) from one signal chain pair at different local sidereal times, combined from 6 nights between April and October 2024. The  ordinate axis scale clips two extreme outliers.  \textit{Right}: Black and gray points show the data from the left panel binned in time (taking the median value of the data in each bin) and multiplied by a frequency-independent calibration coefficient obtained by a linear regression fit of the model sky brightness temperature to the data.  The solid black and gray curves show the model. } 
    \label{fig:galaxy}
\end{figure}

We use the diffuse Galactic synchrotron emission to calibrate the signal chains to determine the amplitude of the electric field of cosmic ray air shower emission measured by the LWA antennas. Figure \ref{fig:galaxy} shows the mean power from both dipoles of a single antenna over different local sidereal times. Events that trigger a data readout and pass the impulsiveness cut provide a sample of the noise background in the first half of the saved timeseries, prior to the triggering event itself. The dataset shown in Figure \ref{fig:galaxy} uses the impulsive events from the six nights of data presented in section \ref{sec:classify}.   The mean power is measured as the mean square of the filtered digital voltage time series.  The power rises and falls as the Galactic plane rises and sets, as well as showing occasional episodes of elevated power attributed to RFI. The right panel of figure \ref{fig:galaxy} shows the sky brightness temperature modeled in \cite{Dowell2017TheSurvey},  integrated over the antenna beam using tools in the LWA Software Library \cite{Dowell2012lsl}, and then integrated over frequency with weights according to the frequency response of the FIR used in the cosmic ray system.  For comparison, the OVRO-LWA mean power data have been binned in time to lower the statistical uncertainty, taking the median value of the data in each time bin in order to reduce sensitivity to episodes of RFI, and then have been converted to brightness temperature by scaling by a calibration factor obtained from a linear regression fit of the model to the data. A separate calibration factor is obtained for each signal chain. The differences between the different antenna's signal chains are likely dominated by the attenuation produced by differing cable lengths and, to a lesser extent, by the impedance matching between each component of the signal chain. The similarity between the scaled power data and the model show that Galactic synchrotron emission dominates the noise power in the OVRO-LWA.  

Calibration coefficients  are sensitive to temperature and other time-varying conditions, whereas Figure \ref{fig:galaxy} combines data from April-October 2024, scaled by one time-independent coefficient in order to demonstrate the response to Galactic synchrotron emission with a wide range of LST coverage.  For the purposes of obtaining calibrated electric fields for cosmic ray pulses, calibration coefficients will be updated more often and will be obtained by the following frequency-dependent approach.

The RMS of component of the ADC output in a frequency range $\nu$ to $\nu + \Delta \nu$ is related to the beam-averaged sky temperature $T_{\rm{model}}$ by
\begin{equation}
    v_{\rm{rms}}(\nu)=|a(\nu)| \frac{\nu}{c} \sqrt{ Z_0 k T_{\rm{model}}\Delta\nu}
\end{equation}
where $Z_0$ is the impedance of free space, $k$ is Boltzmann's constant, and $c$ is the speed of light, and $a(\nu)$ is the calibration coefficient in question.  This result is derived in~\ref{appendix:fieldcal}.

We model $T_{\rm{model}}$ using the LWA Software Library (LSL) \cite{Dowell2012lsl} \texttt{driftcurve} script, which returns a beam-averaged temperature for a given LST and frequency computed from a model sky map\cite{Dowell2017TheSurvey} (expressed in temperature) by projecting this map into azimuth and elevation coordinates at the given LST and then integrating over the above-horizon sky after weighting by the antenna gain in each direction.  The remaining factors are known physical constants and known frequencies, allowing us to solve for $a(\nu)$ by linear regression in 1\,MHz bins.  
The presence of a mesh ground screen under the OVRO-LWA antennas makes the thermal radiation from the ground small, allowing $T_{\rm{model}}$ to be approximated by LSL's integral over the visible sky.

After obtaining $|a(\nu)|$, applying these calibration coefficients and a beam model evaluated for the known arrival direction of the event yields a calibrated timeseries electric field (~\ref{appendix:fieldcal}, equation~\ref{eq_calibratedE}).

The cosmic ray firmware and software have the ability to send a trigger signal on demand (often called a minimum bias trigger), which could be used in the future to regularly read out background noise for calibration purposes.

\section{Acceptance}\label{acceptance}

In this section, we use the instrument model developed in preceding sections and this section and the known cosmic ray flux to predict the expected event rate for the \hbox{OVRO-LWA}. 
The simulations used to estimate the expected event rate are based on the work presented in Monroe et al.~\cite{Monroe2020}. The method of evaluating the event rate and the cosmic ray air shower modeling is the same but with updates to the detector and trigger models based on the instrument developments presented in this paper. 

\subsection{Detector Model}\label{sec:accept.detector}

The antenna effective height is given by 
\begin{equation}
|h_\mathrm{eff}(\lambda, \theta_Z, \phi)|^2 = \frac{4 R_A}{Z_0} \frac{\lambda^2}{4\pi}\frac{|Z_\mathrm{in}|^2}{|Z_A + Z_\mathrm{in}|^2} \mathcal{F}(\lambda)D(\theta_z, \phi)
\end{equation}
where $R_A$ is the antenna radiation resistance, $Z_A$ is the complex antenna impedance, and $Z_\mathrm{in}$ is the front-end input impedance, $Z_0$ is the impedance of free space, $\lambda$ is the wavelength, and $D(\theta_z,\phi)$ is the antenna's directivity.
The factor $\mathcal{F}(\lambda)$ describes the frequency response of the signal chain after the antenna, including bandpass filters applied both in hardware and digitally.

The antenna radiation resistance~$R_A$ and complex impedance~$Z_A$ are unchanged from the descriptions in Monroe et al.~\cite{Monroe2020} (their Figure~23).
The input impedance $Z_\mathrm{in}\simeq100 \,\Omega$.
The factor $\mathcal{F}(\lambda)$ is obtained by comparing the model to measurements of the background noise spectrum (e.g., Figure~\ref{antenna}b).

\begin{figure}
    \centering
    \includegraphics[width=0.8\textwidth, trim={0cm 4cm 0cm 5cm}, clip]{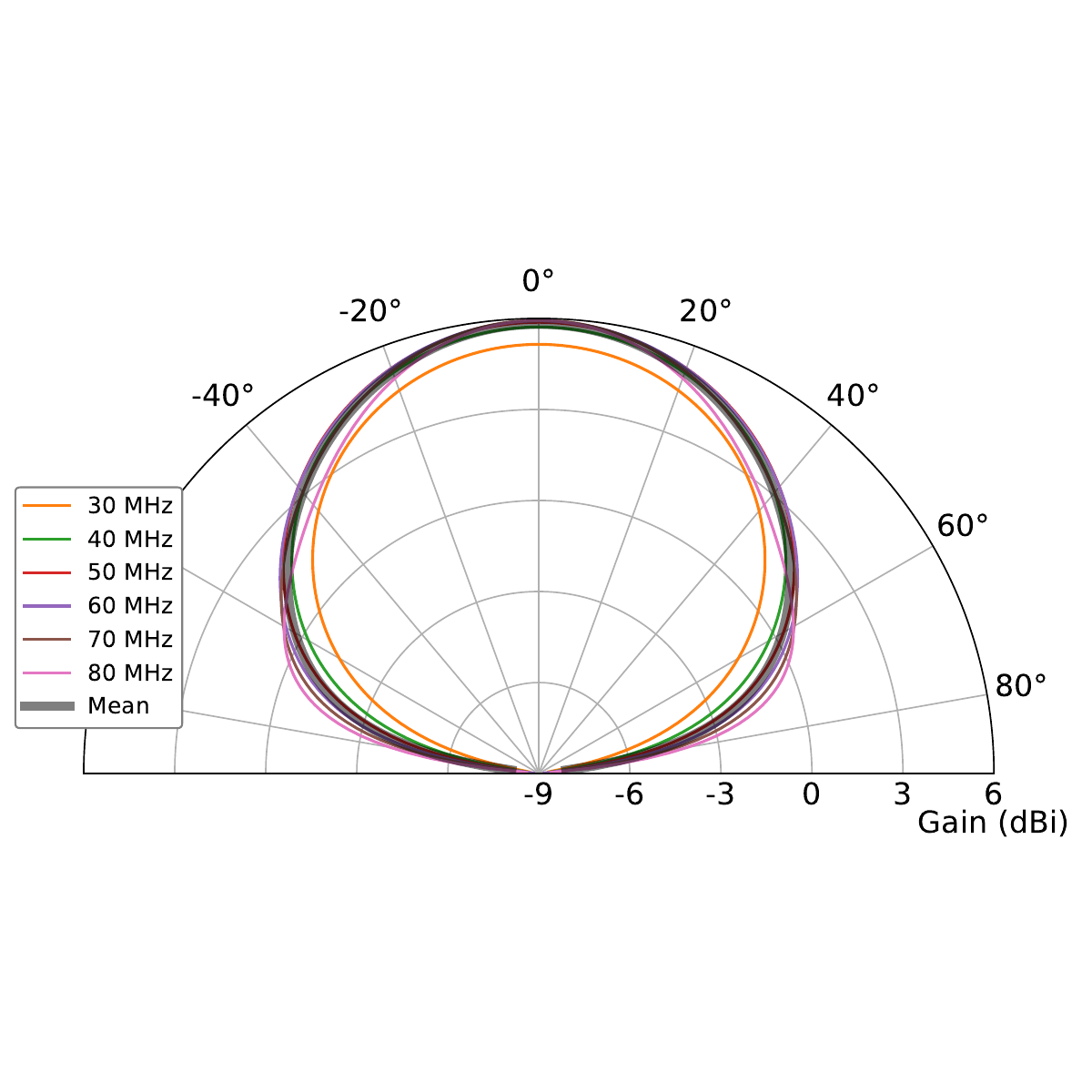}
    \caption{Simulated beam pattern from 30\,MHz to~80\,MHz in steps of~10\,MHz. The mean beam pattern is shown as a thick gray line. }
    \label{beampattern}
\end{figure}

The beam pattern has also been updated from the previous NEC2 model~\cite{Monroe2020} to an updated model using \hbox{FEKO}. Figure~\ref{beampattern} shows the antenna directivity in units of dBi for frequencies between~30\,MHz and~80\,MHz. 
We use the mean of the beam pattern. 
We have also included the azimuthal response of the beam, which is closely approximated by a sine function. 

The noise voltage spectrum~$V_\mathrm{rms}$ at the front end is treated in the same way as~\cite{Monroe2020} using the~\cite{Cane1979} average sky noise spectrum.

\subsection{Trigger Model}\label{sec:accept.trigger}

The trigger model has been updated to match the new antenna FPGA mapping and to include the veto described in Section~\ref{sec:trigger}. 
We model the power signal-to-noise ratio~S/N$_\mathrm{p}$ summed over four time bins as was done in~\cite{Monroe2020} using $\mathrm{S/N_p}\simeq(V_\mathrm{pk}^2 + 4 V_\mathrm{rms}^2)/(4V_\mathrm{rms}^2)$ with $V_\mathrm{pk}$ modeled using $V_\mathrm{pk} = h_\mathrm{eff}E$. Each antenna polarization signal chain triggers on a set threshold crossing and requires $\geq 8$ out of 64 signal chains per FPGA. To set the amplitude signal-to-noise ratio ($\mathrm{S/N_a}$) threshold, we inspected the distribution of $S/N_a$ in each FPGA for the set of 8 cosmic ray candidate events to find the threshold $S/N_a\simeq 8.0$. 

\subsection{Projected Event Rate}

In addition to the trigger, the analysis applies additional quality cuts, described in Section~\ref{sec:classify}. Given that the simulation does not include RFI, we limit the quality cuts applied to simulated data to removing events with elevation angle $<15^\circ$ (zenith angles $>75^\circ$) and a deadtime of $5\%$. We also include cuts associated with the 2D Gaussian fit to the spatial distribution of S/N described in Section~\ref{sec:classify}. The Gaussian cuts require the lateral scale be between 50~m and 500~m, with RMS residuals $<2$ in units of S/N as well as the fitted amplitude of the Gaussian be $<50$ in units of S/N since this saturates antennas near the peak of the distribution. 

\begin{figure}[tb]
    \centering
    \includegraphics[ width=0.8\textwidth]{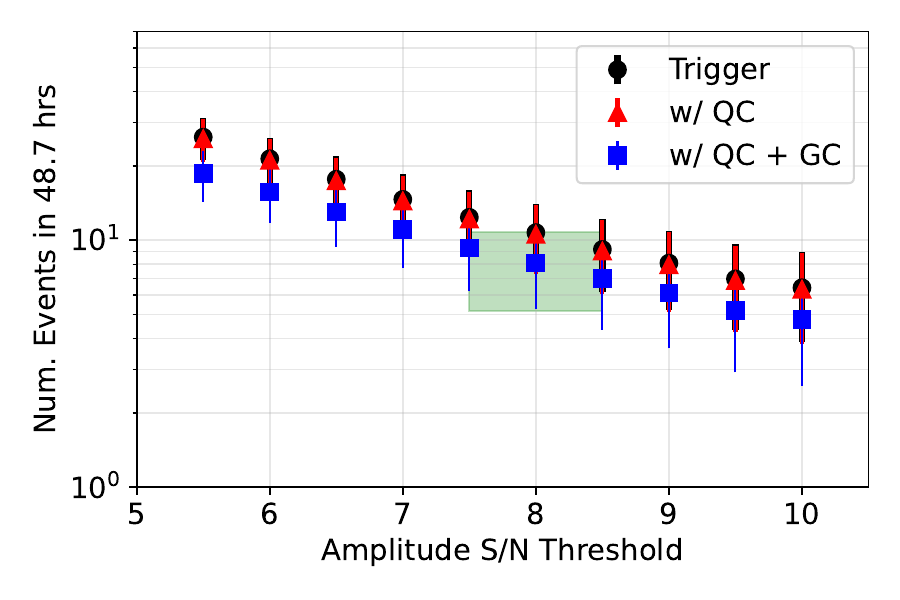}
    \caption{Number of expected events in 48.7 hours of detector uptime as a function of peak amplitude trigger threshold and analysis cuts. The black point (Trigger) show the number of events expected to trigger the array. The error bars include statistical uncertainties only. The red triangles show the event rate after quality cuts (w/ QC) have been applied (see text for details). The blue squares also include Gaussian quality cuts (w/ QC + GC). The light green square highlights the region consistent with 8 events ($\pm\sqrt{8}$ statistical uncertainty) and the empirically-derived amplitude S/N threshold with a $\pm 0.5$ uncertainty to account for S/N fluctuations.  }
    \label{fig:threshold_scan}
\end{figure}

In Figure~\ref{fig:threshold_scan}, we show expected rates based on the trigger model, quality cuts (QC) and Gaussian cuts (QC) described above as a function of amplitude S/N threshold. The light green square indicates the region in the x-axis around our empirically determined threshold S/N of 8.0 with a width of $\pm$0.5 to account for uncertainties in S/N determination and a region around the y-axis corresponding to the 8 events detected with a width of $\pm\sqrt{8}$ to account for Poisson statistical fluctuations.

\begin{figure}[tb]
    \centering
    \includegraphics[ width=0.49\textwidth]{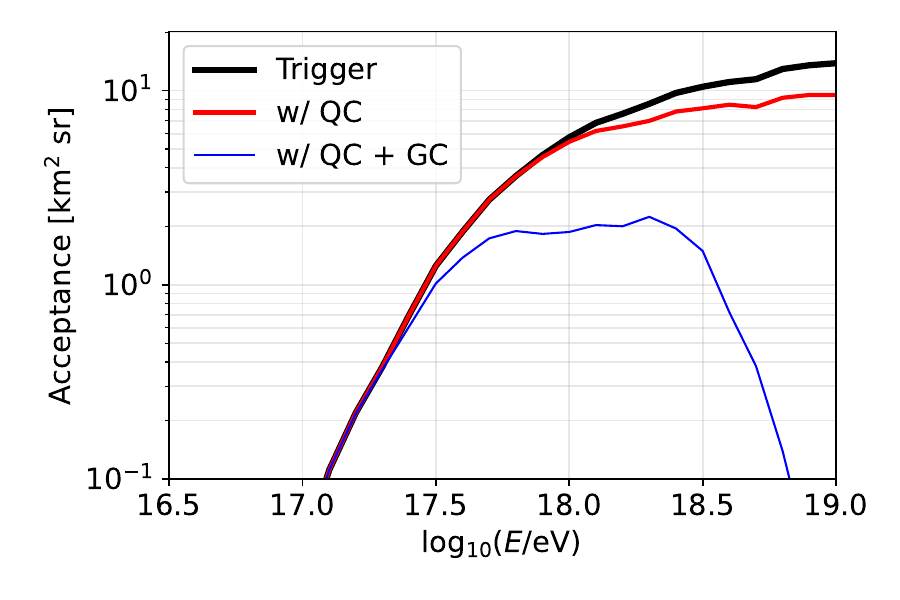}
    \includegraphics[ width=0.49\textwidth]{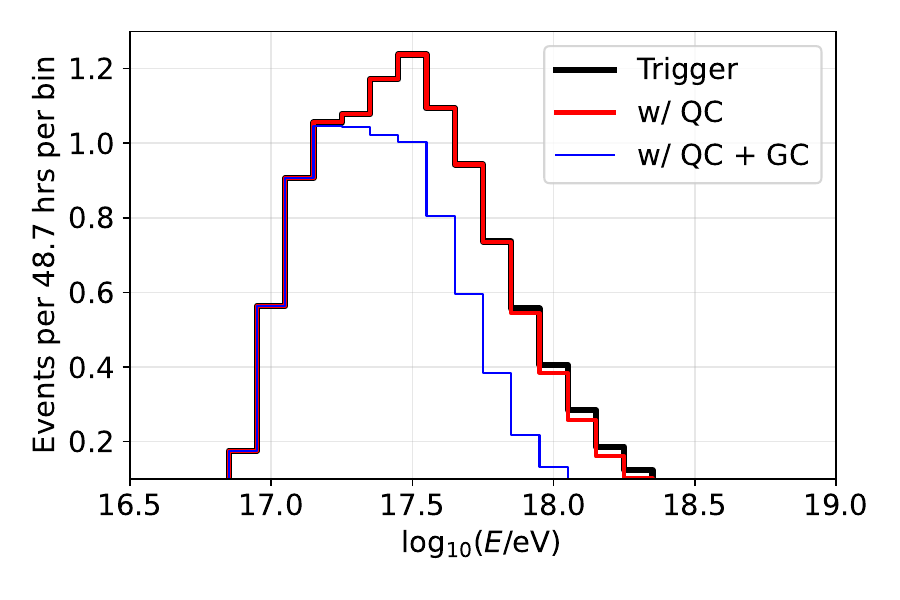}
    \caption{Left: acceptance curves corresponding to an amplitude S/N threshold of 8 (see Figure~\ref{fig:threshold_scan}). The quality cuts and Gaussian cuts reducte the sensitivity to higher energies. Right: the corresponding event rate vs energy distribution. See text for details.}
    \label{fig:acceptance_and_rate}
\end{figure}

In Figure~\ref{fig:acceptance_and_rate} we show the acceptance and event rate vs energy corresponding to an amplitude S/N threshold cut of 8. The quality cuts have a minor effect at energies above $10^{18}$~eV due to the zenith angle cut above $75^\circ$ because at higher energies that radio emission at these high zenith angles is strong enough to trigger the array. The Gaussian cuts have a significant impact above $10^{17.5}$~eV mainly due to the saturation cut requiring a 2D Gaussian amplitude $<50$ in units of S/N.

\subsection{Zenith Angle Distribution}
As an additional consistency check, we show the distribution of zenith angles from simulations and compare them to the population of cosmic-ray candidates. The Gaussian cuts shift the predicted zenith angle distribution towards smaller values since higher zenith angles tend to produce flatter illumination patterns. The cosmic ray candidate zenith angles, shown as green vertical bars, are shown for comparison with the predicted distribution. A comparison of the cumulative distribution function of the cosmic-ray candidate sample and the simulated distribution trigger with quality and Gaussian cut results in a maximum deviation $D=0.216$, which gives a KS test p-value of $p=0.78$. This result indicates consistency between the data and the simulated distribution but the event sample is too small to conclude that they come from the same distribution.

\begin{figure}[tb]
    \centering
    \includegraphics[ width=0.8\textwidth]{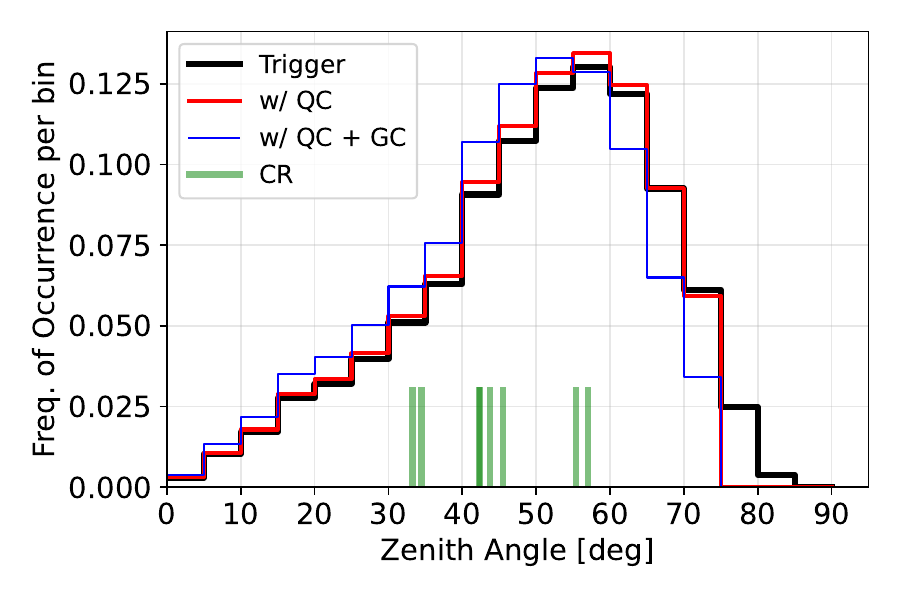}
    \caption{Predicted cosmic-ray zenith angle of arrival for triggered events (Trigger) including quality cuts (w/ QC) and Gaussian cuts (w QC + GC). The observed zenith angle of arrival for the cosmic-ray candidate events (CR) are shown as semi-transparent green vertical bars. Note that two events have angles 42.3$^\circ$ and 42.4$^\circ$, which overlap within the width of one darker green bar. }
    \label{fig:sim_zenith_distrib}
\end{figure}

\subsection{Veto Losses}
One of the main innovations of the OVRO-LWA 352 cosmic ray detector is the implementation of dedicated RFI veto antennas (Section~\ref{sec:fpgaveto}). While these veto antennas are widely separated from the cluster of trigger antennas in each FPGA, a sufficiently strong cosmic ray event could trigger and illuminate veto antennas in such a way as to reject the event. 

\begin{figure}[tb]
    \centering
    \includegraphics[ width=0.8\textwidth]{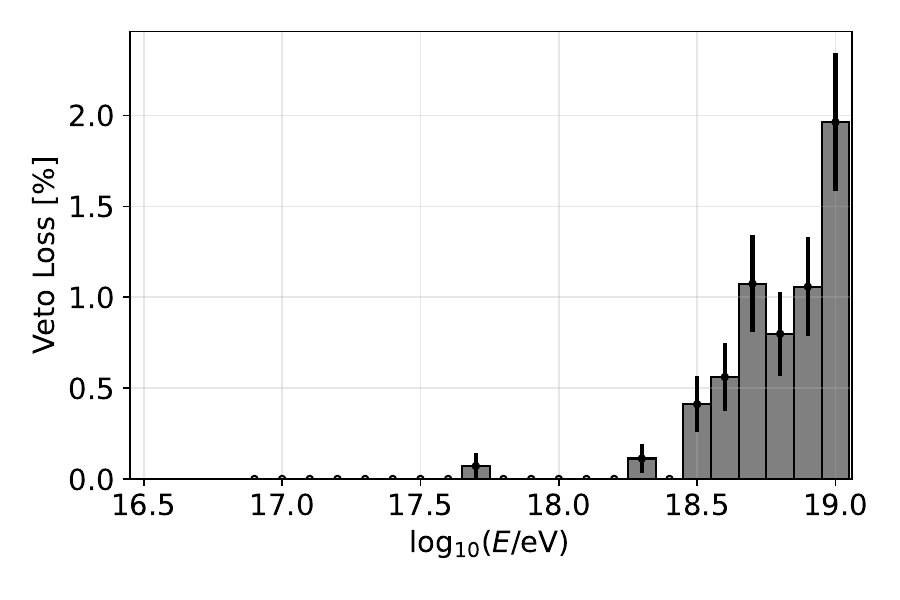}
    \caption{The fraction of simulated cosmic ray events that triggered the array for a trigger with amplitude S/N threshold of 8 but were lost to the veto vs. cosmic ray energy. The error bars are purely statistical from 100,000 events simulated with order-of-magnitude 1,000 triggered per energy bin.}
    \label{fig:veto_loss}
\end{figure}

Figure~\ref{fig:veto_loss} shows the fraction of events that would have triggered the array but were rejected due to the implementation of the veto. The results show that the veto has negligible impact in the energy band of the array of $10^{16.5} - 10^{18.5}$~eV (see right panel of Figure~\ref{fig:acceptance_and_rate}). At higher energies ($>10^{18.5}$~eV), the illumination footprint is sufficiently large to cause some events to be vetoed with the effect growing to $\sim2\%$ at $10^{19}$~eV. 

The implementation of the veto allows the OVRO-LWA to operate as a cosmic ray detector despite the presence of background RFI that would otherwise swamp the detector live time. We have demonstrated that with judicious choice of where the veto antennas are located in relation to the cluster of antennas used for triggering, the array can operate efficiently in its energy band of sensitivity.

\section{Discussion and Future Outlook}\label{sec:discussion}

The OVRO-LWA has detected 8 cosmic rays candidates in 48.7~hours of night time operation with consistent detector settings, 
corresponding to a projected rate of~$480\pm170$ candidates per year for nightly observing.  In addition to the 8 cosmic ray candidates identified during the 48.7~hour dataset with constant detector settings, 6 additional candidates were identified in datasets from other phases of the commissioning process that pass the final selection criteria. These are not used in the rate estimate due to the varying detector sensitivity.
Additional signal attenuation has been added before the digitizer input to prevent daytime RFI from saturating the ADCs, and we plan to explore commissioning the cosmic ray search for operation during daytime RFI conditions.  Furthermore, the impulsivity cut, which was developed as part of the selection cuts that run on the CPUs, could be implemented within the FPGA logic with the potential to reduce the trigger rate by a factor of~40.   On-FPGA impulsivity cuts may be particularly important for small arrays where fewer antennas are illuminated or large, sparse arrays where signals from different antennas cannot be digitized in a central location.

 Radio experiments such as LOFAR reconstruct \xm from the observed cosmic ray air shower radio emission by a method of forward modelling in which a large number of air showers with different \xm are simulated to determine the \xm that best produces the observed radio emission \cite{Corstanje2021a}.  The OVRO-LWA can expect to reconstruct \xm to better than 20 g/cm$^2$ with this method, as modeled in \cite{Carvalho2019}. 

Since the mean and variance of \xm for a sample of air showers are proportional to $-ln \langle a \rangle$, where $\langle a \rangle$ is the mean atomic mass number of the cosmic ray primaries, a large sample of air showers detected with the OVRO-LWA and this state-of-the-art \xm reconstruction approach could be used to make a new, independent estimate of the mass composition of cosmic ray air showers in the 100-1000 PeV energy range.  Under this state of the art, however, the uncertainty in the mass composition is dominated by the systematic uncertainty in the constant of proportionality relating  \xm to $-ln \langle a \rangle$. That uncertainty stems from uncertainties in the hadronic models used to simulate the cosmic ray air showers, which involve collision scenarios beyond what has been observed with the LHC. The state-of-the-art hadronic models disagree over the \xm scale, and none produce a muon shower component consistent with observations \cite{Albrecht2022muon}, leading some recent analyses (e.g. \cite{vicha2025heavyXmax}) to suggest completely shifting the scale of \xm to heavier element interpretations.

The dense antenna array layout will allow the OVRO-LWA to go beyond \xm analyses as a proxy for composition.   Composition-dependent differences between different hadronic interaction models affect not only  \xm but also other details of the shower shape, including a length scale parameter (L). Simulations suggest that simultaneous reconstruction of both L and  \xm by dense radio arrays has the potential to rule out hadronic interaction models and also to provide a precise measurement of the proton fraction in high-energy cosmic rays \cite{Buitink2021}. Independent reconstruction of multiple shower parameters by the forward-modeling method is prohibitively computationally expensive, but the dense antenna spacing of the OVRO-LWA make it the ideal array to test a technique proposed by \cite{schoorlemmer2021interferometry} that applies interferometry to generate three-dimensional tomographic maps of the shower.  This technique is applicable when many antennas thoroughly sample the air shower footprint, as is the case for the OVRO-LWA and the planned 131,072-antenna Square Kilometer Array, making reconstruction techniques tested with the OVRO-LWA applicable to the much larger SKA.

\section{Conclusions}\label{sec:conclusion}

The OVRO-LWA cosmic ray detection system has found 14 cosmic ray candidates, at a rate of one per six hours with optimized detector threshold settings.  The array operates in a complex RFI environment without reference to an external particle trigger. RFI mitigation consists of a combination of RFI rejection in the trigger firmware and in an event analysis software pipeline. Events greater than 15 degrees from the horizon are classified as cosmic rays or RFI using the properties of the radio emission alone, without requiring direction-based source rejection.  The sample of cosmic ray candidates, identified by the shape of their waveforms and the shape of their spatial illumination pattern, has polarization properties consistent with the expectations for cosmic ray emission---without polarization having been consulted as a selection criterion. The array layout of the OVRO-LWA makes it uniquely suited to exploring novel air shower reconstruction techniques for composition studies of cosmic rays at the Galactic to extragalactic transition.

\section{Acknowledgements}
K.P. thanks Andrew Ludwig for a code suggestion that accelerated loading saved data files. The NASA Postoctoral Program and the National Science Foundation Graduate Research Fellowship (under Grant No. 2139433) have supported K.P.'s work.
A portion of this research was carried out at the Jet Propulsion Laboratory, California Institute of Technology, under a contract with the National Aeronautics and Space Administration (80NM0018D0004).  This material is based in part upon work supported by the National Science Foundation under grants number AST-952 1828784 and PHY-2310134, the Simons Foundation (668346, JPG), the Wilf Family Foundation and Mt.~Cuba Astronomical Foundation. We thank the Xilinx University Program for donating software licenses used in compiling FPGA firmware.

\appendix
\section{ADC Synchronization and digital data time-stamping}\label{appendix:timing}
\begin{figure}[tb]
    \centering
    \includegraphics[ width=0.6\textwidth]{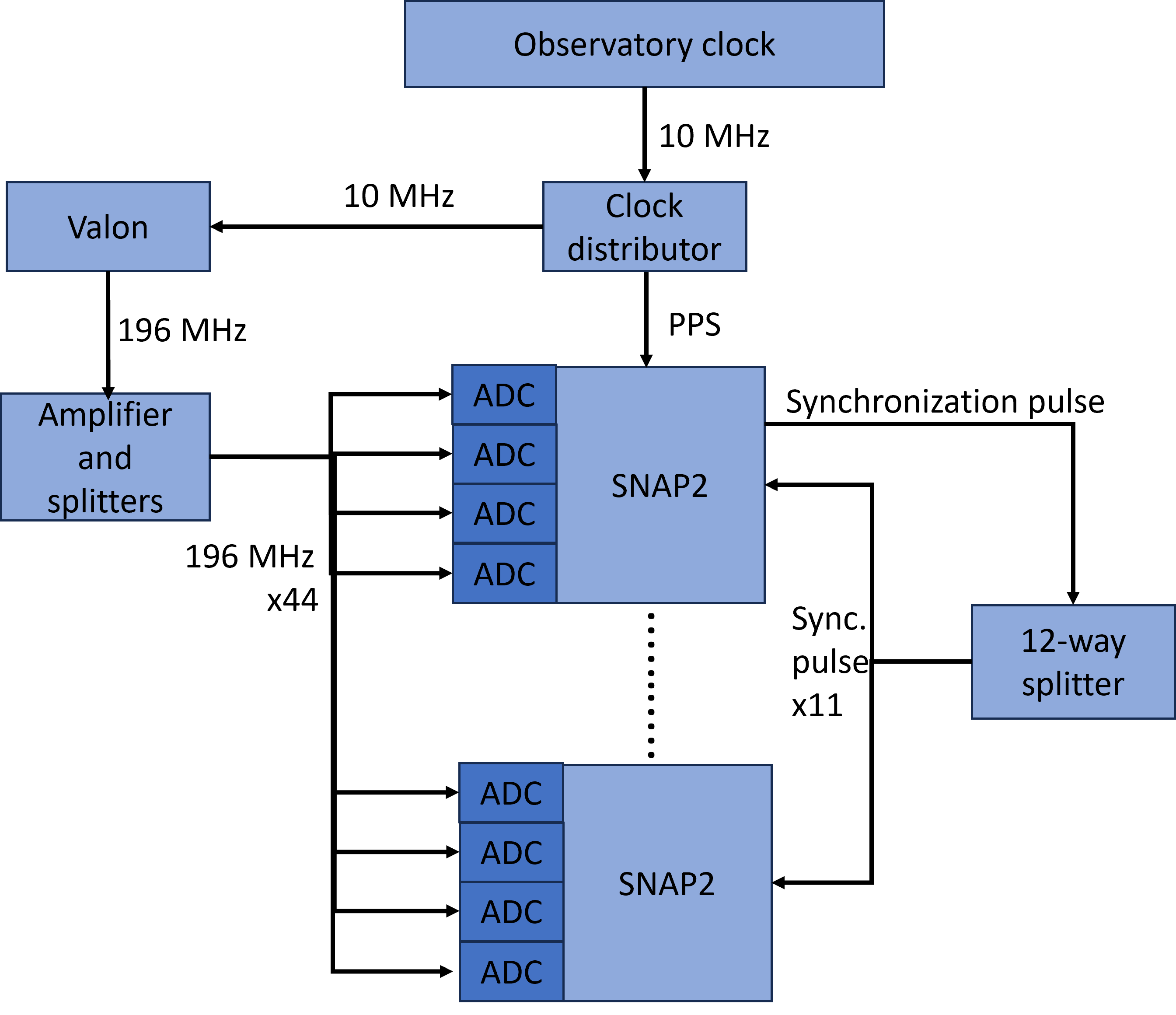}
    \caption{Clock distribution system for the \hbox{OVRO-LWA}. A 10\,MHz signal from the main observatory clock also transmits a 1\,Hz signal by missing one pulse every second. The clock distributor outputs a PPS signal and a 10\,MHz reference separately.  The Valon signal generator outputs a 196\,MHz sinusoid that is split to all of the 44 ADCs cards using 44 cables of equal length (to better than one centimeter). The PPS signal is sent to one SNAP2 board, which distributes a synchronization pulse to all the others and back to itself via a splitter, again with all 11 cables the same length.}
     \label{fig:lwa-clock}
\end{figure}

 The synchronization and timestamping for the digital signal processing system is accomplished as follows (Figure~\ref{fig:lwa-clock}). A coaxial cable from the main observatory clock carries a 10\,MHz reference signal, which includes a 1\,Hz signal by missing one pulse every second. A clock device in the LWA signal processing shelter takes that observatory clock signal and outputs a Pulse-Per-Second (PPS) signal and a 10\,MHz reference separately. A Valon signal generator locks to the 10\,MHz signal and outputs a 196\,MHz sinusoid that is split to all of the 44 ADC cards using 44 cables of the same length (to within better than a clock cycle delay).  The clock that drives the FPGAs on the SNAP2s is taken from the ADC cards. 

In order to timestamp the data, the pulse-per-second signal is input to one SNAP2 board, which distributes a single-clock-cycle synchronization pulse to the other 10 boards and back to itself, via a splitter (Figure \ref{fig:lwa-clock}).  All the output cables from the splitter are the same length to better than one centimeter. The synchronization pulse repeats with a known period greater than one second, which is set from software on start up.  On startup, the first SNAP2 board receives a 64-bit integer indicating the UTC time of the next PPS pulse, expressed in number of clock cycles since the Unix epoch. 
When each of the other SNAP2 boards are synchronized, which need not be performed simultaneously for all the boards, they receive the 64-bit integer timestamp that will correspond to the next pulse from the SNAP2 that is distributing the synchronization signal. When that pulse arrives, it triggers a synchronization pulse that propagates through the FPGA logic to every firmware component that uses a time reference. Within each of these subsystems, on receipt of the synchronization pulse, a counter begins counting up from the received Unix time value, incrementing by one on each clock cycle. The output of these counters can be used to timestamp the data in each firmware subsystem. The values of these 64-bit integer timestamps from all the SNAP2 boards thus refer to one time reference, from which the relative time of events at different antennas can be determined.  When cosmic ray data is read out into ethernet packets, this 64-bit timestamp is included.

 \section{Derivation of the Electric field Amplitude Calibration }\label{appendix:fieldcal}

 We derive the calibration approach by first explaining the relation between the measured signal and the incident electric field from a point source, and then show how the necessary calibration terms can be obtained by comparing measurements of the background noise to a model of the Galactic synchrotron emission.

An electric field incident on an antenna will produce an electric field across the antenna's terminals with 
\begin{equation}
    v_a = \textbf{E} \cdot \textbf{h} = Eh\cos{\alpha} = E_\parallel h
\end{equation}

where $\textbf{h}$ is the effective height of the antenna, $h=|\textbf{h}|$, $E=|\textbf{E}|$, and $\alpha$ is the angle between the polarization of electric field vector and the antenna effective height, and $E_\parallel$ is the component of the electric field parallel to h. In this analysis we neglect polarization cross terms, which would replace the vector $\textbf{h}$ with a tensor.  

The goal of the calibration is to relate the discrete timeseries output from the ADCs, $v_{\rm{d}}[n]$, to a sampled electric field timeseries $E_\parallel [n]$.  Since $h$ and various system gains are frequency dependent, we consider the discrete Fourier transforms of these timeseries. Let $E_\parallel[k]$ and $v_{\rm{d}}[k]$ be the components of the discrete Fourier transforms (DFTs) of $E_\parallel [n]$ and $v_{\rm{d}}[n]$, respectively.

Let the angular dependence of the antenna response be described by $g(\theta,\phi,\nu) =\frac{h(\theta,\phi,\nu)^2}{h_{\rm{max}}(\nu)^2} $ where $h_{\rm{max}}(\nu)^2$ is the maximum value of the effective height over all directions, at the frequency $\nu$, for a wave arriving from direction zenith angle $\theta$ and azimuth $\phi$ . Then

\begin{equation}
v_a[k] = E_{\parallel}[k]h_{\rm{max}}(\nu[k]) \sqrt{g(\theta,\phi,\nu)}
\end{equation}

where $\nu[k]$ is the frequency corresponding to the $k^{th}$ component of the DFT.

The incident electric field from a cosmic ray shower is superposed on a stochastic electric field due to Galactic synchrotron emission. This emission is the largest source of noise in the system. The voltage at the input to the ADC can be approximated as
\begin{equation}\label{eq_vadc}
v_{\rm{ADC}}[k] = f_a(\nu[k]) ( v_{a}[k] + s[k])
\end{equation}
where $f(\nu[k])$ are the frequency-dependent amplitude gain factors due to the signal path from the antenna to the ADC and $s[k]$ are the DFT components of the system noise. In an exact treatment, each of the elements in the analog signal path described in Section 2 contribute a gain factor, and each of the active elements contribute additional noise terms, but Equation \ref{eq_vadc} is a good approximation when the noise due to the Galactic synchrotron emission is much larger than the noise due to the electronics, as is the case for the OVRO-LWA.    

The ADC adds a frequency independent digital gain, $f_d$, such that the components of the DFT of the ADC output are $v_{d}[k] = f_d v_{ADC}[k] $.

Thus, if the DFT components of the electric field at times can be estimated from the ADC output, to within the uncertainty set by the noise term, as:  
\begin{equation}\label{eq_E}
    E_\parallel[k] = \left(f_d f(\nu[k]) h_{\rm{max}}(\nu[k]) \sqrt{g(\theta,\phi, \nu)} \right)^{-1} v_d[k] \pm s[k] 
\end{equation}

By taking the inverse DFT, the time series of the electric field can be estimated---to within uncertainty equal to the square root of the variance of $s$ for Gaussian distributed noise---as:
\begin{equation}
    E_\parallel[n] = \frac{1}{N}\sum_{k=0}^{N-1}\left(f_d f_a(\nu[k]) h_{\rm{max}}(\nu[k]) \sqrt{g(\theta,\phi, \nu[k]}) \right)^{-1} v_d[k] e^{i 2 \pi n k/N}
\end{equation}

\begin{equation}\label{eq_calibratedE}
    = \frac{1}{N}\sum_{k=0}^{N-1}\left( a(\nu[k]) \sqrt{g(\theta,\phi, \nu[k]}) \right)^{-1} v_d[k] e^{i 2 \pi n k/N}
\end{equation}
where $a(\nu)$ = $f_d f_a(\nu[k]) h_{\rm{max}}(\nu[k])$.  Note that $f$ and $g$ are dimensionless, $h_{\rm{max}}$ has dimensions of length, $v_{\rm{d}}$ has ADC units, and $f_a$ has dimensions of voltage per ADC unit.

We obtain $g(\theta,\phi, \nu)$ from simulations of the antennas, and obtain the calibration coefficients $a(\nu)$ by relating the measured rms voltage ADC output to the known brightness temperature of Galactic synchrotron emission as follows.

Assume $a(\nu)$ is smooth such that it can be approximated as a piecewise function with a constant value in frequency bands of width $\Delta \nu$, with $\Delta \nu$ larger than the frequency resolution of the DFT.  The RMS of the components of the ADC output and the electric field in a frequency range from $\nu$ to $\nu + \Delta \nu$ are 
\begin{equation}
    v_{\rm{rms}}^2(\nu) = \sum_{k=k_1}^{k=k_2} |v_d[k]|^2 , ~ ~ ~ ~ E_{\rm{rms}}^2 = \sum_{k=k_1}^{k=k_2} |E[k]|^2
\end{equation}
where $k_1$ and $k_2$ are the indices of the Fourier coefficients corresponding to $\nu$ and $\nu + \Delta \nu$, respectively.  
Since $a(\nu)$ is roughly constant across this range, 
\begin{equation}
    v_{\rm{rms}}^2=a(\nu)g(\theta,\phi,\nu) \mathrm{RMS}[cos(\alpha)E]^2,
\end{equation}
which, for unpolarized Galactic synchrotron emission, is 
\begin{equation}
    v_{\rm{rms}}^2(\nu) = \frac{a(\nu)}{2} g(\theta,\phi,\nu) E_{\rm{rms}}^2
\end{equation}

Galactic emission from the whole sky contributes to the RMS voltage at the ADCs, and so if emission from direction $(\theta,\phi)$ contributes $E'^2_{\rm{rms}}(\theta,\phi,\nu) \sin(\theta) d\theta d\phi$ to the mean squared electric field in frequency band  ($\nu$, $\nu + \Delta \nu$) then the mean square ADC output in that band is 
\begin{equation}
    v_{\rm{rms}}^2(\nu) = \frac{a(\nu)}{2}\int_0^{2\pi} \int_{0}^{\pi/2} E'^2_{\rm{rms}}(\theta,\phi,\nu) g(\theta\phi\nu) \sin(\theta) d\theta d\phi
\end{equation}

 The Poynting  flux,  $F$, delivered by an electromagnetic wave is related to the RMS of the radiated electric field by $F = \frac{E^2_{\rm{rms}}}{Z_o}$, where $Z_o$ is the impedance of free space
\footnote{Note that the Poynting flux $F=\frac{E^2_{pk}}{2Z_o} = \frac{E^2_{\rm{rms}}}{Z_o}$ where $E^2_{pk}$ is the amplitude of the incident wave}.
The brightness temperature, $T_{\rm{sky}}(\theta,\phi,\nu)$, is defined such that the specific intensity (flux per unit solid angle per unit frequency) $I = \frac{2 \nu^2}{c^2}k T_{\rm{sky}}$ where $k$ is Boltzmann's constant. $I \Delta \nu$ is the total flux per solid angle incident on the antenna in the band of width $\Delta \nu$, and this is equal to the Poynting flux per unit solid angle of the incident radiation.  Thus,
\begin{equation}
    E'^2_{\rm{rms}}(\theta,\phi,\nu) = Z_o I(\theta,\phi,\nu) \Delta \nu = \frac{2 Z_o \nu^2}{c^2}k T_{\rm{sky}}\Delta \nu
\end{equation}

\begin{equation}
    v^2_{\rm{rms}} = \frac{a(\nu)}{2} \int_0^{2\pi} \int_{0}^{\pi/2} \frac{2 Z_o \nu^2}{c^2}k T_{\rm{sky}}\Delta \nu g(\theta,\phi,\nu) \sin(\theta)d\theta d\phi 
    =  a(\nu)Z_0 k \frac{\nu ^2}{c^2} \Delta \nu T_{\rm{model}}
\end{equation}

Thus,
\begin{equation}
T_{\rm{model}} \equiv \int_0^{2\pi} \int_{0}^{\pi/2} T_{\rm{sky}} g(\theta,\phi,\nu) \sin(\theta)d\theta d\phi
\end{equation}
\begin{equation}
    v_{\rm{rms}}(\nu)=|a(\nu)| \frac{\nu}{c} \sqrt{ Z_0 k\Delta \nu T_{\rm{model}}}
\end{equation}

Note that working with frequency bins wider than the frequency resolution allows the estimate of the RMS voltage to approach a Gaussian-distributed estimator by the central limit theorem, whereas the amplitudes $v_d[k]$ should be Rayleigh distributed.  We use 1\,MHz bins.



\input{output2.bbl}\end{document}

%% file: CR-instrument-paper-authors.tex
  \author[ a, b,1,2]{Kathryn A.~Plant, \note{Corresponding Author.}\note{Nasa Postdoctoral Program Fellow, currently Jansky Fellow at NRAO}}

  \author[ a]{Andrew Romero-Wolf,}  

\author[ c, b]{Gregg Hallinan,}

\author[ a, b]{Marin M.~Anderson,}

\author[ d]{Judd D.~Bowman,}

\author[ c,3]{Ruby Byrne,\note{NSF Postdoctoral Fellow}}

\author[ e]{Bin Chen,}

\author[ e]{Xingyao Chen,}

\author[ b]{Morgan Catha,}

\author[ e, f]{Sherry Chhabra,}

\author[ c, b]{Larry D’Addario,}

\author[ c, b, g]{Ivey Davis,}

\author[ h]{Jayce Dowell,}

\author[ d]{Katherine Elder,}

\author[ e]{Dale Gary,}
\author[ b]{Charlie Harnach,}

\author[ c, b]{Greg Hellbourg,}

\author[i]{Jack Hickish,}

\author[ b]{Rick Hobbs,}

\author[ c]{David Hodge,}

\author[ b]{Mark Hodges,}

\author[ c, b]{Yuping Huang,}

\author[j,k]{Andrea Isella,}

\author[ d]{Daniel C.~Jacobs,}

\author[ b]{Ghislain Kemby,}

\author[ b]{John T.~Klinefelter,}

\author[ d]{Matthew Kolopanis,}

\author[ c, b]{Nikita Kosogorov,}

\author[ b]{James Lamb,}

\author[ c, b]{Casey Law,}

\author[ c, b,4]{Nivedita Mahesh,\note{David and Ellen Lee Postdoctoral Fellow}}

\author[ e]{Surajit Mondal,}

\author[ e]{Brian O’Donnell,}

\author[ b]{Corey Posner,}

\author[ b]{Travis Powell,}

\author[ b]{Vinand Prayag,}

\author[ b]{Andres Rizo,}

\author[ d]{Jun Shi,}

\author[ h]{Greg Taylor,}

\author[ b]{Jordan Trim,}

\author[ b]{Mike Virgin,}

\author[ d]{Akshatha Vydula,}

\author[ c]{Sandy Weinreb,}

\author[ b]{Scott White,}

\author[ b]{David Woody,}

\author[ e]{Sijie Yu,}

\author[ b]{Thomas Zentmeyer,}

\author[ e,l]{Peijin Zhang,}

\author[ a,5]{T.~Joseph~W.~Lazio,\note{Current affiliation: University of Michigan}\note{\copyright 2025. All rights reserved}}

\affiliation[ a]{
   Jet Propulsion Laboratory, California Institute of
    Technology,
    4800~Oak Grove~Dr,  
   Pasadena,
   CA, 
   91109, 
   USA}

\affiliation[ b]{Owens Valley Radio Observatory, California Institute
    of Technology,\\ 100 Leighton Lane, Big Pine,CA,93513,USA}

\affiliation[ c]{
   Cahill Center for  Astronomy and Astrophysics,
    California Institute of Technology,
    1200~E~California~Blvd, 
   Pasadena, 
   CA, 
   91125, 
   USA}

\affiliation[ d]{
   School of Earth and Space Exploration, Arizona State
    University, 
   781~E~Terrace~Rd, 
   Tempe, 
   AZ, 
   85287, 
   USA}

\affiliation[ e]{
   Center for Solar-Terrestrial Research, New Jersey
    Institute of Technology,
    323~Dr~Martin Luther King~Jr Boulevard, 
   Newark, 
   NJ, 
   07102-1982, 
   USA}

\affiliation[ f]{
   George Mason University, 
   4400~University~Dr, 
   Fairfax, 
   VA, 
   22030, 
   USA}

\affiliation[ g]{
       Astron,
       Oude Hoogeveensedijk~4, 
       Dwingeloo, 
       7991~PD, 
       The Netherlands}

\affiliation[ h]{
   University of New Mexico, 
   2500~Campus Blvd~NE, 
   Albuquerque, 
   NM, 
   87131, 
   USA}

\affiliation[i]{
   Real-Time Radio Systems, Ltd, 
   \\Bournemouth,
   Dorset, 
   BH6~3LU, 
   UK}

\affiliation[j]{
   Department of Physics and  Astronomy, Rice University, 6100~Main~St, 
   Houston, 
   TX, 
   77005, 
   USA}

\affiliation[k]{
   Rice Space Institute, Rice University, 6100~Main~St,
   Houston, 
   TX, 
   77005, 
   USA}

\affiliation[l]{
   Cooperative Programs for the Advancement of Earth System Science, University Corporation for Atmospheric Research, 3090~Center Green~Dr, 
   Boulder, 
   CO, 
   80301, 
   USA}
  
\emailAdd{kplant@nrao.edu}